\documentclass[
twocolumn,
superscriptaddress,
 amsmath,amssymb,
 aps,
 pra,
floatfix,
]{revtex4-2}

\usepackage{physics}
\usepackage{graphicx}
\usepackage[breaklinks=true,colorlinks,citecolor=blue,linkcolor=blue,urlcolor=blue]{hyperref}
\usepackage[capitalise]{cleveref}
\usepackage[caption=false]{subfig}
\usepackage{orcidlink}

\newcommand{\Mavg}{\overline{M}_2}
\newcommand{\MH}{M_2^{\mathrm H}}

\begin{document}

\title{Magic in quantum trajectory unravelings of noisy random quantum circuits}

\author{Sara Santos}
\email{sara.alvesdossantos@epfl.ch}
\affiliation{Institute of Physics, Ecole Polytechnique Fédérale de Lausanne (EPFL), CH-1015 Lausanne, Switzerland}
\affiliation{Center for Quantum Science and Engineering, Ecole Polytechnique Fédérale de Lausanne (EPFL), CH-1015 Lausanne, Switzerland}
\author{Alessio Paviglianiti~\orcidlink{0000-0002-2719-7080}}
\email{alessio.paviglianiti@epfl.ch}
\affiliation{Institute of Physics, Ecole Polytechnique Fédérale de Lausanne (EPFL), CH-1015 Lausanne, Switzerland}
\affiliation{Center for Quantum Science and Engineering, Ecole Polytechnique Fédérale de Lausanne (EPFL), CH-1015 Lausanne, Switzerland}
\author{Vincenzo Savona}
\email{vincenzo.savona@epfl.ch}
\affiliation{Institute of Physics, Ecole Polytechnique Fédérale de Lausanne (EPFL), CH-1015 Lausanne, Switzerland}
\affiliation{Center for Quantum Science and Engineering, Ecole Polytechnique Fédérale de Lausanne (EPFL), CH-1015 Lausanne, Switzerland}

\date{\today}

\begin{abstract}
The quantum-trajectory description of a noisy quantum circuit is not unique: every Kraus representation of the noise channel defines a different unraveling, i.e., a different ensemble of pure states that averages to the same density matrix. While all unravelings share the same expectation values of observables, they generally differ in the amount of quantum resources carried by the individual trajectories. Here we study non-stabilizerness, or magic, quantified by the stabilizer R\'enyi entropy, in quantum-trajectory unravelings of Haar-random and $T$-doped random Clifford brickwork circuits subject to local dephasing noise of strength $p$. The canonical unraveling in terms of Pauli jumps is Clifford and leaves magic untouched, so that trajectories retain the Haar-random value of non-stabilizerness. Exploring the full two-parameter family of minimal unravelings, we find that the trajectory-averaged magic can be substantially reduced by a state-independent choice of Kraus operators, and that the optimal choice is not in general the projective-measurement unraveling, but a two-outcome generalized measurement in which the Kraus operators correspond to a weak measurement of the noisy qubit. The optimal unraveling reduces the extensive magic density of the trajectories by a $p$-dependent factor without altering its linear scaling with system size. At the same time, it produces a very broad distribution of magic over trajectories, with a sizable fraction of trajectories retaining Haar-random magic. Since the cost of stabilizer-based classical simulation grows exponentially with the magic of the state, these high-magic trajectories, rather than the average, control the simulation cost. Our results delineate the potential and the limitations of unraveling optimization as a route to the classical simulation of noisy quantum dynamics with stabilizer methods.
\end{abstract}

\maketitle

\section{Introduction}

Quantum information theory provides a powerful toolbox for the investigation of many-body systems and quantum circuits. Quantum resources~\cite{chitambar2019quantumresourcetheories} are now routinely adopted as diagnostics of a wide range of phenomena, driven by the successful application of entanglement~\cite{amico2008entanglemententropy} in the study of many-body effects such as scrambling~\cite{nahum2017quantumentanglementgrowth,keyserlingk2018operatorhydrodynamics,fischer2013randomquantumcircuits}, topological order~\cite{kitaev2006topologicalentanglement,levin2006detectingtopologicalorder}, and quantum phase transitions~\cite{calabrese2004entanglemententropy,laflorencie2016quantumentanglementincondensed}. In recent years, non-stabilizerness (or \textit{magic})~\cite{bravyi2005universalquantumcomputation,bravyi2012magicstatedistillation,veitch2014resourcetheoryofstabilizer,howard2017applicationofaresourcetheory} has rapidly emerged as a key complementary probe, largely thanks to the introduction of computable measures such as the stabilizer R\'enyi entropy~\cite{leone2022stabilizerrenyientropy,haug2023stabilizerentropiesmonotones,leone2024stabilizerentropiesaremonotones} and of efficient numerical estimators for many-body states~\cite{haug2023quantifyingnonstabilizerness,lami2023nonstabilizernessviaperfectsampling,tarabunga2024nonstabilizernessviamps}. This has stimulated intense research on its dynamical behavior in random circuits and chaotic systems~\cite{turkeshi2025magicspreading,niroula2024phasetransitioninmagic,szombathy2025asymptoticallyindependent,paviglianiti2025estimatingnonstabilizerness,russomanno2025nonstabilizernessintheunitary,falcao2025nonstabilizernessdynamics,paviglianiti2026truecostfactoringlinking}, in monitored dynamics~\cite{bejan2024dynamicalmagictransitions,fux2024entanglementmagicseparation,tirrito2025magicphasetransitionsmonitored}, as well as on its connection with chaos~\cite{leone2021quantumchaosis} and quantum criticality~\cite{tarabunga2023manybodymagic,haug2023quantifyingnonstabilizerness}. Fermionic non-Gaussianity is currently undergoing a parallel development~\cite{sierant2026fermionicmagicresources,collura2026nonstabilizernessof,aditya2026growthandspreading}. These advances address the fundamental question of how different quantum resources contribute to making many-body quantum physics complex~\cite{leone2021quantumchaosis,paviglianiti2026emergenceofgenericentanglement}. These resources also carry an operational meaning linked to the cost of classical simulation: entanglement limits the performance of matrix-product state (MPS) techniques~\cite{vidal2003efficientclassicalsimulation,schollwock2011thedensitymatrixrenormalizationgroup}, whereas magic controls the computational overhead of stabilizer expansion methods~\cite{gottesman1998heisenbergrepresentation,aaronson2004improvedsimulation,bravyi2016improvedclassicalsimulation,bravyi2019simulationofquantum,seddon2021quantifyingquantumspeedups,bittel2026operationalinterpretation}.

The resource perspective has lately gained significant interest in the study of many-body open quantum systems, especially driven by the discovery of measurement-induced phase transitions~\cite{li2018quantumzenoeffect,skinner2019measurementinduced,gullans2020dynamicalpurification,turkeshi2021measurementinducedentanglement,jian2020measurementinducedcriticality,paviglianiti2023multipartiteentanglement,agrawal2022entanglementandchargesharpening}. In this context, quantum complexity measures are evaluated at the level of individual realizations of the monitoring protocol, i.e., of quantum trajectories~\cite{dalibard1992wavefunctionapproach,dum1992montecarlosimulation,carmichael1993opensystemsapproach,wiseman1996quantumtrajectories,plenio1998quantumjumpapproach,daley2014quantumtrajectories}, and their ensemble averages exhibit critical behavior invisible in the averaged density matrix~\cite{fischer2013randomquantumcircuits,potter2022entanglementdynamicshybrid}. Understanding how such processes impact quantum complexity contributes to delineating a boundary between quantum and classical regimes: sufficiently strong noise can counteract the scrambling of the unitary dynamics~\cite{noh2020efficientclassicalsimulation,aharonov2023polynomialtimeclassicalalgorithm}, thus potentially making the evolution of trajectories efficiently simulable with classical techniques~\cite{zihan2023efficientsaqmpling,dowling2026noiseinducedsimulabilitytransitionoperator}.

Motivated by these ideas, recent works have explored how to reduce entanglement in the trajectories of open quantum systems in order to enhance the efficiency of MPS simulations. A given dissipative channel admits an infinite number of unravelings, which correspond to ensembles of stochastic realizations that average to a common density matrix. Since quantum complexity measures depend on individual pure states rather than on the ensemble's density matrix, their averages differ across unravelings and can be minimized by an appropriate choice~\cite{vovk2022entanglementoptimaltrajectories,kolodrubetz2023optimalityoflindbladunfolding,chen2024optimizedtrajectoryunraveling,vovk2024quantumtrajectoryentanglement,daraban2025nonunitaritymaximizing,cichy2026classicalsimulationofnoisy}, a freedom that has also been exploited to extend the reach of classical sampling algorithms for noisy circuits~\cite{zihan2023efficientsaqmpling}. However, it has also been pointed out that a lower average entanglement does not automatically translate into a lower simulation cost, as the speed of statistical convergence may differ across unravelings~\cite{sander2026computationalregimesmatrixproductstatebasedquantum}. These efforts have mostly targeted entanglement, because MPS methods are the most established computational framework for many-body systems, and only recently has the unraveling been optimized for simulations combining Clifford circuits and MPS~\cite{fioroni2026noiseinduced}. The question of how magic and other resources behave across different unravelings, as well as how this impacts the efficiency of other classical simulation algorithms, stands open. It is a particularly natural question for stabilizer-based methods, because the canonical unraveling of a Pauli noise channel consists of Pauli operators, which are Clifford: in a stabilizer simulation, such noise is free~\cite{gottesman1998heisenbergrepresentation,aaronson2004improvedsimulation,gidney2021stim}, but for the same reason it cannot remove any of the non-stabilizerness generated by the ideal circuit.

In this work, we address this gap by investigating non-stabilizerness in different unravelings of dissipative dynamics. We consider one-dimensional Haar-random and $T$-doped random Clifford circuits perturbed by local dephasing noise, and we compare the trajectory-averaged stabilizer R\'enyi entropy across the full two-parameter family of minimal unravelings of the channel, as well as the nonminimal unraveling in terms of projective measurements. We show that the ``canonical'' unraveling is sub-optimal with respect to the resource: in Haar-random circuits, the representation of the noise process in terms of Pauli jumps is unable to change non-stabilizerness at all, and trajectories retain the magic of Haar-random states. As our main result, we demonstrate that much better choices can be made, even in an agnostic way that does not rely on information on the current state to adapt the unraveling dynamically. Contrary to the intuition suggested by the literature on monitored circuits, the optimal unraveling in the high-magic regime is not the one in terms of projective measurements, but rather corresponds to a generalized measurement; the projective unraveling prevails only when magic is scarce, as in weakly $T$-doped Clifford circuits. We quantify the magic reduction by showing that, while the scaling remains extensive with the system size, its prefactor is lowered by the optimal protocol. Finally, we observe that while the average resource can be reduced, its distribution over trajectories becomes very broad, and a considerable number of typical trajectories retain near-maximal magic. Stabilizer-expansion methods are thus limited by these high-magic trajectories rather than by the mean. These are, however, sampled considerably less often than in the canonical unraveling, which indicates a more efficient trajectory-based simulation, even though magic remains extensive.

The rest of the manuscript is structured as follows. Section~\ref{sec:theory} defines the noisy random circuit setup under study, and presents the theoretical tools and numerical methods: Sec.~\ref{sec:noise} introduces the noise channels and the parameterization of their unravelings, Sec.~\ref{sec:trajectories} covers the quantum-trajectory formalism, while Sec.~\ref{sec:magic} recalls the definition of the stabilizer R\'enyi entropy and the algorithm used to evaluate it exactly. Section~\ref{sec:results} presents our numerical results for the magic dynamics, its dependence on the unraveling parameters in Haar-random and $T$-doped Clifford circuits, its scaling with system size, and its distribution over trajectories. Section~\ref{sec:discussion} summarizes our findings and discusses the implications of these results from the perspectives of resource theory and stabilizer-based classical simulation, also outlining open directions.

\section{Theory and methods}
\label{sec:theory}

\begin{figure}[t]
  \includegraphics[width=\columnwidth]{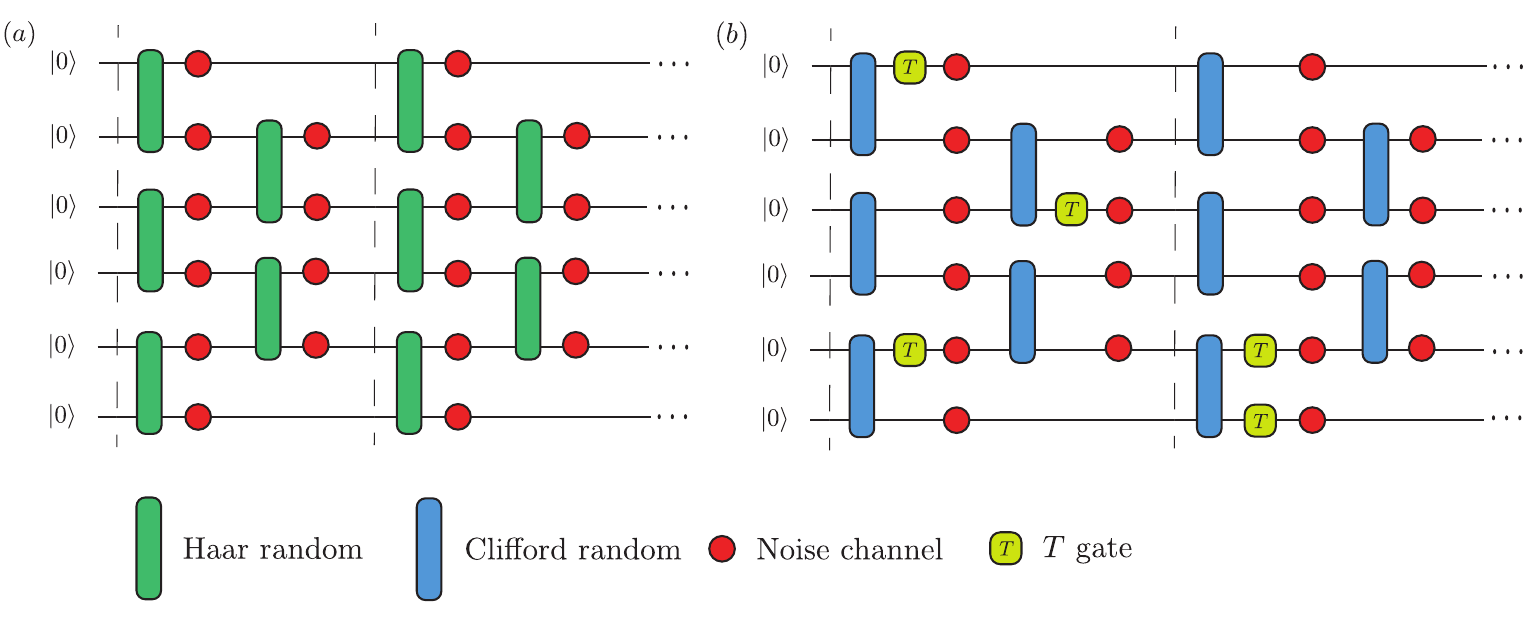}
  \caption{\textbf{Random quantum-circuit ensembles.} All qubits are
  initialized in $\ket{0}$. The vertical dashed lines delimit complete
  brickwork layers, each formed by two sublayers acting on complementary
  sets of nearest-neighbor bonds. (a) Haar-random circuit, in which each
  green block is an independently sampled two-qubit Haar unitary.
  (b) $T$-doped Clifford circuit, in which each blue block is an
  independently sampled two-qubit Clifford and a $T$ gate is applied
  independently to each participating qubit with probability $q$. Red
  circles denote the local single-qubit noise channel applied after each
  ideal gate block.}
  \label{fig1}
\end{figure}

We consider a system of $n$ qubits, with Hilbert space of dimension $d=2^n$, whose state is described by a density operator $\rho$. The system evolves under a noisy quantum circuit, in which every ideal two-qubit gate is followed by the independent action of a single-qubit noise channel on each of the two qubits involved in the gate. We consider noisy random quantum circuits with the one-dimensional brickwork architecture depicted in Fig.~\ref{fig1}: a layer consists of two consecutive sublayers of two-qubit gates, one acting on each of the two complementary sets of disjoint nearest-neighbor bonds. Within this circuit geometry, we focus on two classes of gates, illustrated in Fig.~\ref{fig1}(a) and (b). In Haar-random circuits, every two-qubit gate is drawn independently from the Haar measure on $\mathrm{U}(4)$~\cite{mezzadri2007howtogenerate}; in the absence of noise, these circuits generate magic rapidly within the first few layers~\cite{turkeshi2025magicspreading}. In $T$-doped Clifford circuits~\cite{zhou2020singletgate,leone2021quantumchaosis,haferkamp2023efficientunitarydesigns}, this rate can instead be tuned. Each two-qubit gate is a uniformly random Clifford unitary~\cite{koenig2014howtoefficientlyselect,bravyi2021hadamardfreecircuits}, and each of the two qubits then receives a $T$ gate, $T=\mathrm{diag}(1,e^{i\pi/4})$, with probability $q$. The doping probability $q$ thus controls how fast magic is generated: at $q=0$ the circuits are purely Clifford and generate no magic, while larger values of $q$ inject magic faster, although even at $q=1$ the circuits remain distinct from Haar-random ones.

All qubits are initialized in the stabilizer product state
$\ket{0}^{\otimes n}$, i.e., $\rho_0=(\ket{0}\!\bra{0})^{\otimes n}$,
so that any non-stabilizerness present at later times is generated by
the circuit. Denoting by $\Lambda_\ell$ the completely positive,
trace-preserving map that describes the $\ell$-th layer, gates and
noise included, the state after $N_L$ layers is
\begin{equation}
    \rho
    =
    \Lambda_{N_L}\circ\cdots\circ\Lambda_{2}\circ\Lambda_{1}(\rho_0).
    \label{eq:rho_evolution}
\end{equation}
In the absence of noise, $\rho$ remains pure at all times. In the
presence of noise, $\rho$ is a mixed state, and it can be represented
as a statistical mixture of pure states in infinitely many ways \cite{wiseman_inequivalence_2001,wiseman1996quantumtrajectories,wiseman2010quantummeasurementandcontrol}. The
central theme of this work is that each such representation can be \textit{unraveled} into a different ensemble of pure-state quantum
trajectories, and that a nonlinear property of the trajectories, such
as their non-stabilizerness, depends on the unraveling although $\rho$
does not.

\subsection{Noise models and unravelings}
\label{sec:noise}

We describe noise by local, Markovian quantum channels~\cite{breuer2002theoryofopenquantumsystems}. A channel
$\mathcal{E}$ acting on a single-qubit density matrix admits a Kraus
representation
\begin{equation}
    \mathcal{E}(\rho)
    =
    \sum_{\mu=0}^{r-1}
    K_{\mu}\rho K_{\mu}^{\dagger},
    \qquad
    \sum_{\mu=0}^{r-1}
    K_{\mu}^{\dagger}K_{\mu}=I ,
    \label{eq:kraus_representation}
\end{equation}
where the second relation ensures trace preservation. Here and below, $I$
denotes the identity operator on one qubit, and $\rho$ in
Eq.~\eqref{eq:kraus_representation} may equally denote the reduced
state of the qubit or the full $n$-qubit state, on which $K_\mu$ acts
as $K_\mu\otimes I^{\otimes(n-1)}$. As stated above, the channel acts
independently on both qubits of a gate immediately after the gate
as illustrated in Fig.~\ref{fig1}.

The Kraus representation of a channel is not unique. Given a set of $r$
Kraus operators $\{K_\mu\}$, let $V$ be an $s\times r$ isometry, with $V^\dagger V=I_r$.
The operators
\begin{equation}
    \widetilde K_a
    =
    \sum_{\mu=0}^{r-1}V_{a\mu}K_\mu,
    \qquad a=0,\ldots,s-1,
    \label{eq:kraus_isometry}
\end{equation}
define the same channel, since
\begin{equation}
    \sum_{a=0}^{s-1}
    \widetilde K_a\rho\widetilde K_a^\dagger
    =
    \sum_{\mu=0}^{r-1}
    K_\mu\rho K_\mu^\dagger .
\end{equation}
For $s=r$, the transformation is unitary and preserves the number of
Kraus operators. A rectangular isometry with $s>r$ instead produces a
nonminimal representation with a larger number of outcomes. While all
such representations give identical density-matrix dynamics, they
generally define different ensembles of quantum trajectories. We refer
to each choice of Kraus representation as an unraveling of the
channel~\cite{carmichael1993opensystemsapproach,gisin1992quantumstatediffusion,wiseman1996quantumtrajectories,wiseman2010quantummeasurementandcontrol}.

In the main text we focus on the dephasing, or phase-flip, channel;
the analogous constructions for depolarizing and amplitude-damping
noise, together with the corresponding numerical results, are given in
Appendix~\ref{app:noise}. The dephasing channel reads
\begin{equation}
    \mathcal{D}_p(\rho)
    =
    (1-p)\rho+pZ\rho Z ,
    \label{eq:dephasing_channel}
\end{equation}
whose canonical Kraus operators are
\begin{equation}
    K_0=\sqrt{1-p}\,I,
    \qquad
    K_1=\sqrt{p}\,Z .
    \label{eq:canonical_dephasing}
\end{equation}
The off-diagonal entries of $\rho$ are multiplied by $1-2p$.
We therefore focus on $0\leq p\leq 1/2$, for which increasing $p$
corresponds to increasing dephasing strength, with complete dephasing
at $p=1/2$.

Up to phases multiplying the individual Kraus operators, any minimal
unraveling of Eq.~\eqref{eq:dephasing_channel} can be
parameterized by two angles. We use
\begin{equation}
    U(\theta,\phi)
    =
    \begin{pmatrix}
        \cos\theta & -\sin\theta\\
        \sin\theta &  \cos\theta
    \end{pmatrix}
    \begin{pmatrix}
        e^{i\phi} & 0\\
        0 & e^{-i\phi}
    \end{pmatrix},
    \label{eq:unraveling_unitary}
\end{equation}
where one may take $0\leq\theta\leq\pi/2$ and
$0\leq\phi<\pi$. The transformed Kraus operators are defined by
\begin{align}
    K_0^{(\theta,\phi)}
    &=
    e^{i\phi}\sqrt{1-p}\cos\theta\,I
    -
    e^{-i\phi}\sqrt{p}\sin\theta\,Z,
    \nonumber\\
    K_1^{(\theta,\phi)}
    &=
    e^{i\phi}\sqrt{1-p}\sin\theta\,I
    +
    e^{-i\phi}\sqrt{p}\cos\theta\,Z .
    \label{eq:dephasing_unraveling_explicit}
\end{align}
The canonical unraveling is recovered at $\theta=0$, since the
remaining phases do not affect either the outcome probabilities or the
conditional pure states. Note that both Kraus operators of the
canonical unraveling are proportional to Clifford unitaries ($I$ and
$Z$). Any nonlinear function of the state that is invariant under
Clifford operations, and in particular the stabilizer R\'enyi entropy
introduced below, is therefore unaffected by the noise events of the
canonical unraveling: in this representation, dephasing noise can
neither create nor destroy magic along a trajectory. For $\theta\neq
0$, instead, the Kraus operators are non-unitary and generally
non-Clifford, and the noise events do modify the magic of the state.

For comparison, we also consider an unraveling associated with a
projective measurement on the $Z$ eigenbasis. Introducing the projectors onto the eigenstates of $Z$,
\begin{equation}
    P_\pm=\frac{I\pm Z}{2},
\end{equation}
and applying a $3\times2$ isometry, as in Eq.~\eqref{eq:kraus_isometry}, to the canonical Kraus operators, we can rewrite the channel as
\begin{equation}
\mathcal{D}_p(\rho) = (1-2p)\rho + 2p\left(P_+\rho P_+ + P_-\rho P_-\right),
\label{eq:dephasing_measurement_decomposition}
\end{equation}
which corresponds to the three Kraus operators
\begin{equation}
K_{\mathrm{id}}=\sqrt{1-2p}\,I, \quad K_+=\sqrt{2p}\,P_+, \quad K_-=\sqrt{2p}\,P_- .
\label{eq:projdeph}
\end{equation}
In this unraveling, with probability $1-2p$ the state is left unchanged; otherwise, the qubit is measured in the computational basis and projected onto $\ket{0}$ or $\ket{1}$. The noise thus acts as a measurement in the computational basis, performed at random with probability $2p$. Since the probability $1-2p$ of the identity outcome must be nonnegative, this representation is valid only for $0\leq p\leq 1/2$, the range already considered above. At $p=1/2$, the identity outcome disappears and every noise event becomes a measurement.

Within the minimal family of Eq.~\eqref{eq:dephasing_unraveling_explicit}, a distinguished member is obtained at $\phi=0$ and
\begin{equation}
    \theta^\ast=\arctan\sqrt{\frac{p}{1-p}}.
    \label{eq:thetastar}
\end{equation}
Inserting this value in Eq.~\eqref{eq:dephasing_unraveling_explicit} and using $I=P_++P_-$, $Z=P_+-P_-$, one finds
\begin{equation}
    K_0^{\ast}=(1-p)I-pZ=I-2p\,P_+,
    \quad
    K_1^{\ast}=2\sqrt{p(1-p)}\,P_+ .
    \label{eq:thetastar_kraus}
\end{equation}
This is a quantum-jump unraveling: the ``click'' operator $K_1^\ast$ is proportional to the projector onto the $+1$ eigenstate of $Z$ and occurs with probability $2p(1-p)(1+\langle Z\rangle)$, whereas the ``no-click'' operator $K_0^\ast$ is a weak measurement of strength $2p$ that leaves the $P_-$ component untouched and slightly suppresses the $P_+$ component. Comparing Eq.~\eqref{eq:thetastar_kraus} with Eq.~\eqref{eq:projdeph}, one finds
\begin{equation}
    K_0^\ast=\sqrt{1-2p}\,K_{\mathrm{id}}+\sqrt{2p}\,K_-,
    \qquad
    K_1^\ast=\sqrt{2(1-p)}\,K_+ .
\end{equation}
The $\theta^\ast$ unraveling thus corresponds to the projective one with the identity and $P_-$ outcomes merged coherently into a single Kraus operator. For a given state, this merging changes nonlinear quantities such as $M_2$ at first order in $p$. In Appendix~\ref{app:thetastar}, however, we show that in Haar-random circuits these first-order differences cancel on average, so that the trajectory-averaged $M_2$ of the two unravelings differs only at order $p^2$. We will see below that the $\theta^\ast$ unraveling yields a strongly reduced magic with respect to the canonical one, but that it is not the optimal choice within the family of Eq.~\eqref{eq:dephasing_unraveling_explicit}.

\subsection{Quantum trajectories}
\label{sec:trajectories}

Consider a system in a pure state $\ket{\psi}$. As discussed in Sec.~\ref{sec:noise}, the action of a noise channel on this state can be written through its Kraus decomposition, Eq.~\eqref{eq:kraus_representation},
\begin{equation}
    \mathcal{E}\big(\ket{\psi}\!\bra{\psi}\big) = \sum_{\mu=0}^{r-1} K_\mu\ket{\psi}\!\bra{\psi}K_\mu^\dagger .
    \label{eq:channel_on_pure_state}
\end{equation}
Each term in this sum is proportional to a pure state. Defining the outcome probabilities and the corresponding normalized states,
\begin{equation}
    \pi_\mu=\bra{\psi}K_\mu^\dagger K_\mu\ket{\psi}, \qquad \ket{\psi_\mu}=\frac{K_\mu\ket{\psi}}{\sqrt{\pi_\mu}} ,
\end{equation}
Eq.~\eqref{eq:channel_on_pure_state} becomes
\begin{equation}
    \mathcal{E}\big(\ket{\psi}\!\bra{\psi}\big) = \sum_{\mu=0}^{r-1} \pi_\mu\ket{\psi_\mu}\!\bra{\psi_\mu} = \mathbb{E}_{\mu}\big[\ket{\psi_\mu}\!\bra{\psi_\mu}\big],
    \label{eq:channel_trajectory_decomposition}
\end{equation}
where $\mathbb{E}_{\mu}$ denotes the average over the random outcome $\mu$, drawn with probability $\pi_\mu$. The output of the channel is thus a statistical mixture of pure states. This observation underlies the quantum-trajectory method~\cite{dalibard1992wavefunctionapproach,dum1992montecarlosimulation,carmichael1993opensystemsapproach,plenio1998quantumjumpapproach,daley2014quantumtrajectories}, in which the channel is replaced by a stochastic update: $K_\mu$ is applied with probability $\pi_\mu$, and the state is then renormalized. By Eq.~\eqref{eq:channel_trajectory_decomposition}, this update reproduces the channel exactly on average. Repeating it at every noise event generates a pure-state trajectory $\ket{\psi_\gamma}$, labeled by the sequence of outcomes $\gamma=(\mu_1,\mu_2,\ldots)$.

In a Monte Carlo calculation with $N_t$ trajectories, this average is estimated as
\begin{equation}
    \rho
    \simeq
    \frac{1}{N_t}
    \sum_{\gamma=1}^{N_t}
    \ket{\psi_\gamma}\!\bra{\psi_\gamma},
    \label{eq:trajectory_monte_carlo}
\end{equation}
with statistical fluctuations that decrease asymptotically as
$1/\sqrt{N_t}$.

Different unravelings of the same channel reproduce the same density
matrix, and therefore the same
expectation values for observables linear in $\rho$. They nevertheless
produce different probability distributions over pure states.
Consequently, the trajectory average of a nonlinear pure-state
quantity $\mathcal{Q}$,
\begin{equation}
    \overline{\mathcal{Q}}_{\{K_\mu\}}
    =
    \mathbb{E}_{\gamma}
    \left[
        \mathcal{Q}\!\left(\ket{\psi_\gamma}\right)
    \right],
    \label{eq:nonlinear_trajectory_average}
\end{equation}
generally depends on the chosen Kraus representation. This dependence
is central to the present work. While the mixed-state evolution is
unraveling-independent, the magic carried by the individual
trajectories is not.

\subsection{Stabilizer R\'enyi entropy}
\label{sec:magic}

To quantify magic of a pure $n$-qubit state $|\psi\rangle$, we
use the second stabilizer R\'enyi entropy
(SRE)~\cite{leone2022stabilizerrenyientropy}, defined as
\begin{equation}
    M_2(|\psi\rangle)
    =
    -\ln\left[
        \frac{1}{d}
        \sum_{P\in\mathcal{P}_n}
        \left|
        \langle\psi|P|\psi\rangle
        \right|^4
    \right],
    \label{eq:M2_definition}
\end{equation}
where $\mathcal{P}_n=\{I,X,Y,Z\}^{\otimes n}$ is the set of $4^n$ Pauli strings. 
The SRE thus depends on the fourth moment of the distribution of Pauli expectation values. It vanishes if and only if $\ket{\psi}$ is a stabilizer state, is invariant under Clifford unitaries, which makes it a convenient measure of the non-Clifford resources contained in a state~\cite{leone2022stabilizerrenyientropy,haug2023stabilizerentropiesmonotones,leone2024stabilizerentropiesaremonotones}. For Haar-random states
the SRE concentrates around the value~\cite{leone2022stabilizerrenyientropy,turkeshi2025paulispectrum}
\begin{equation}
    \MH
    =
    \ln\!\left(\frac{d+3}{4}\right)
    \simeq (n-2)\ln 2 ,
    \label{eq:haar_magic}
\end{equation}
which is the maximal amount of magic that a typical state can host up
to an $\mathcal{O}(1)$ correction, and which grows extensively with
$n$. We shall use $\MH$ as the reference value against which the
magic of noisy trajectories is compared. 

Evaluating Eq.~\eqref{eq:M2_definition} directly, however, is computationally demanding. It requires all $4^n$ Pauli expectation values and,
for a generic state vector, has an overall computational complexity
of order $\mathcal{O}(8^n)$~\cite{Huang2026XORFWHT}. To reduce this cost, we evaluate Eq.~\eqref{eq:M2_definition} exactly using the XOR–fast Walsh–Hadamard transform (XOR–FWHT) method~\cite{Huang2026XORFWHT}. 

The total computational cost of evaluating $M_2$ with this method scales as $\mathcal{O}(n4^n)$. This represents a substantial improvement over the $\mathcal{O}(8^n)$ scaling of the direct evaluation while remaining deterministic and exact. Nevertheless, the cost is still exponential in the number of qubits, so exact evaluations of $M_2$ remain restricted to comparatively small or medium system sizes. Since $M_2$ must be evaluated for a large number of trajectories, at every circuit layer and over a fine grid of unraveling parameters, we restrict ourselves to system sizes for which its exact evaluation remains affordable ($n\leq12$). Stochastic estimators of the SRE, such as perfect Pauli sampling of matrix product states~\cite{lami2023nonstabilizernessviaperfectsampling,haug2023quantifyingnonstabilizerness,tarabunga2024nonstabilizernessviamps}, could extend the analysis to larger systems in regimes of moderate entanglement, at the price of an additional statistical error.

\section{Results}
\label{sec:results}

We present here numerical results characterizing magic for different unravelings. 
Unless otherwise stated, the results shown in this section refer to $n=8$
qubits and to dephasing noise unraveled according to
Eq.~\eqref{eq:dephasing_unraveling_explicit} or to the projective
unraveling of Eq.~\eqref{eq:projdeph}. A scaling analysis up to $n=12$
is presented in Sec.~\ref{sec:scaling}. We denote by $\Mavg$ the SRE averaged over $N_t$ trajectories for each of $N_c$ independent circuit realizations, and we mostly report it normalized by the Haar-random value $\MH$ of Eq.~\eqref{eq:haar_magic}, i.e., as $\Mavg/\MH$.
Statistical error bars are smaller than the symbol size in all SRE-averaged quantities presented.

\begin{figure}[!h]
  \centering
  \includegraphics[width=\columnwidth]{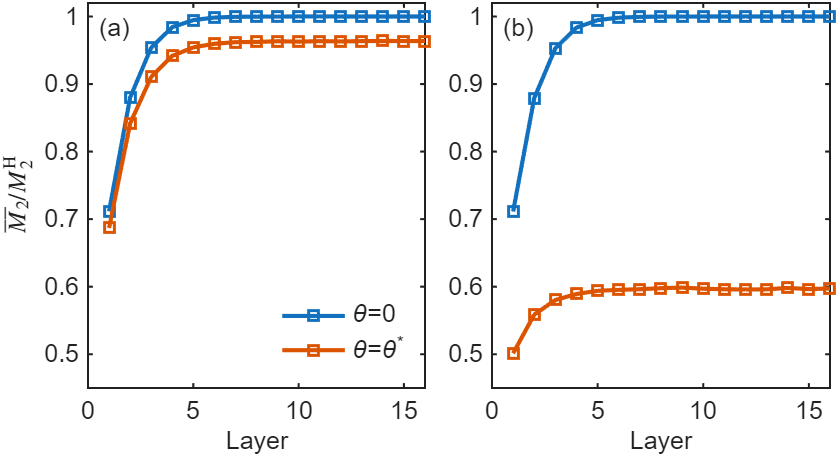}
  \caption{\textbf{Magic dynamics in noisy Haar-random brickwork circuits.}
$\Mavg/\MH$ after each circuit layer, for dephasing probabilities (a) $p=0.01$ and (b) $p=0.1$. Each panel compares the canonical unraveling ($\theta=0$) with the quantum-jump unraveling ($\theta=\theta^\ast$), with $\phi=0$ in both cases. Circuits contain $n=8$ qubits and $16$ layers. Each curve averages over $80$ random circuit realizations, with $1000$ trajectories per realization.}
  \label{fig2}
\end{figure}

\begin{figure}[!h]
  \centering
  \includegraphics[width=\columnwidth]{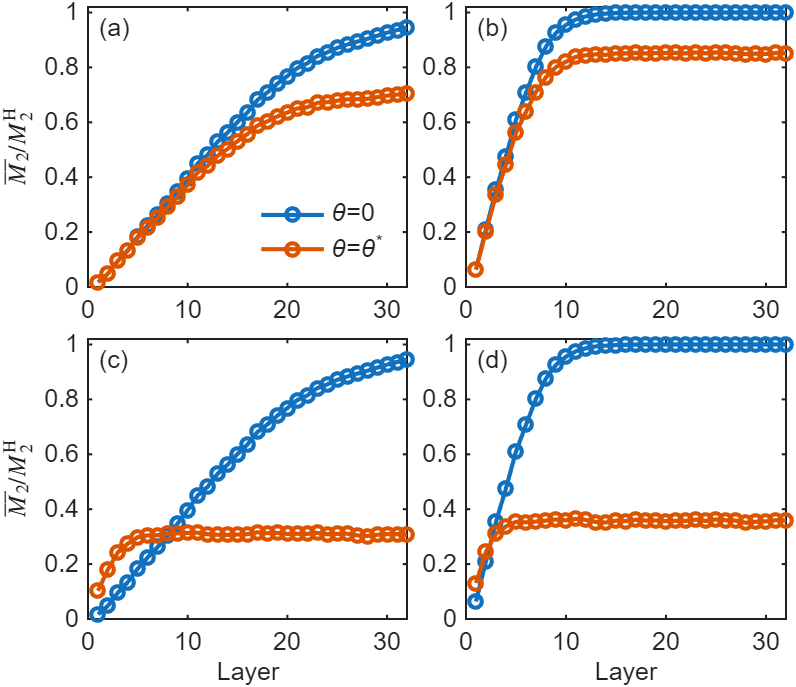}
  \caption{\textbf{Magic dynamics in noisy $T$-doped random Clifford brickwork circuits.}
$\Mavg/\MH$ after each circuit layer, for (a) $p=0.01$, $q=0.05$; (b) $p=0.01$, $q=0.2$; (c) $p=0.1$, $q=0.05$; and (d) $p=0.1$, $q=0.2$. Each panel compares the canonical unraveling ($\theta=0$) with the quantum-jump unraveling ($\theta=\theta^\ast$), with $\phi=0$ in both cases. Circuits contain $n=8$ qubits and $32$ layers. Each curve averages over $80$ random circuit realizations, with $1000$ trajectories per realization.}
  \label{fig3}
\end{figure}

\subsection{Magic dynamics}
\label{sec:dynamics}

Figs.~\ref{fig2} and \ref{fig3} show the trajectory-averaged SRE
evaluated after each layer of the circuit, for the Haar-random and
$T$-doped Clifford ensembles respectively. The results are plotted for the canonical unraveling at $\theta=0$, and for the quantum-jump unraveling at $\theta=\theta^\ast$. In all cases, magic grows
monotonically with the number of layers and saturates to a limiting
value that depends on the noise strength $p$ and on the unraveling. For
the canonical unraveling, this limiting value is the
Haar-random one, $\Mavg=\MH$, irrespective of $p$, as the one-qubit Clifford Kraus operators leave both the Haar-random and Clifford-random ensembles unchanged~\cite{turkeshi2025magicspreading}. For the
quantum-jump unraveling at $\theta=\theta^\ast$, instead, the
saturation value decreases as $p$ increases. This suppression is a result of the Kraus operators being non-unitary.

In Haar-random circuits, magic is generated rapidly and saturates after a number of layers of the order of the
system size~\cite{turkeshi2025magicspreading}. In what follows, we take $N_L=2n$ layers for Haar-random circuits, well beyond saturation. In $T$-doped Clifford circuits, the growth rate of magic is instead controlled by the doping probability $q$ (Fig.~\ref{fig3}) and can become very small: at $q=0.05$, the canonical unraveling has not yet saturated after $32$ layers for $n=8$. For these circuits, we therefore consider deeper circuits of at least $N_L=4n$ layers, and $N_L=48$ at the lowest doping $q=0.05$, ensuring that saturation is reached. Since $\Mavg$ has then reached its saturation value, we characterize each unraveling by its value at the end of the circuit.

\subsection{Haar-random circuits: dependence on the unraveling}
\label{sec:haar}

\begin{figure*}[t]
  \centering
  \includegraphics[width=0.9\textwidth]{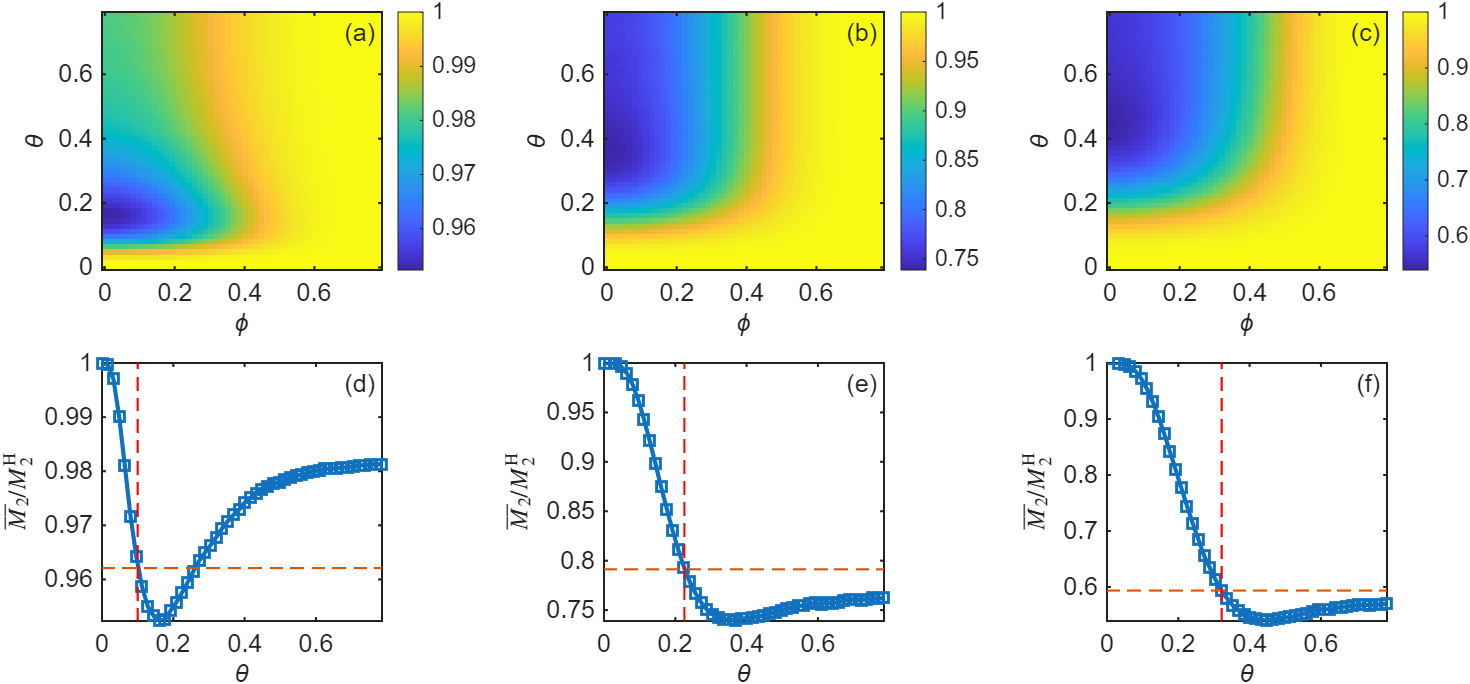}
  \caption{\textbf{Magic in noisy Haar-random brickwork circuits as a function of the unraveling parameters.}
(a)--(c) $\Mavg/\MH$ at the end of circuits with $n=8$ qubits and $N_L=16$ layers, for dephasing probabilities $p=0.01$ in (a),(d), $p=0.05$ in (b),(e), and $p=0.1$ in (c),(f). Panels (a)--(c) sample $50$ values of each unraveling angle, $\theta$ and $\phi$; panels (d)--(f) show the corresponding $\phi=0$ cuts (symbols connected by solid lines). Vertical red dashed lines mark the quantum-jump unraveling, $\theta=\theta^\ast$. Horizontal dashed lines show the values obtained with the projective unraveling of Eq.~\eqref{eq:projdeph}. All averages use $8$ random circuit realizations, with $1000$ trajectories per realization.}
  \label{fig4}
\end{figure*}

Figure~\ref{fig4} shows the trajectory-averaged magic at the end of noisy Haar-random circuits as a function of the unraveling parameters $\theta$ and $\phi$, for three values of the dephasing strength $p$. Two features stand out. First, for any $\theta$, the magic is minimal at $\phi=0$ and increases monotonically with $\phi$ up to $\phi=\pi/4$, where it reaches the Haar value. At this angle, both Kraus operators are proportional to unitary phase gates, which are absorbed by the subsequent Haar-random gates, so the trajectories remain Haar-distributed. Second, at $\phi=0$, the magic depends non-monotonically on $\theta$: it starts from the Haar value at $\theta=0$, decreases steeply, reaches a minimum at a finite angle $\theta_{\min}$, and then increases again towards $\theta=\pi/4$. Larger values of $\theta$ need not be considered: the replacement $\theta\to\pi/2-\theta$ maps the Kraus pair $(K_0,K_1)$ onto $(K_1,K_0)$ up to the sign of the $Z$ term, i.e., onto the mirror unraveling obtained by conjugation with the bit flip $X$. Since the circuit ensembles are invariant under a bit flip on the noisy qubit (see Appendix~\ref{app:thetastar}) and $M_2$ is invariant under Clifford operations, the average magic is symmetric about $\theta=\pi/4$. Similarly, $\phi$ and $\phi+\pi/2$ are related by a sign flip of the $Z$ term and yield the same average magic, while $\phi$ and $-\phi$ are related by complex conjugation, which also leaves $M_2$ and the Haar ensemble invariant. It is therefore sufficient to consider $0\leq\phi\leq\pi/4$.

The role of the phase $\phi$ can be understood from the structure of the Kraus operators, which are diagonal in the computational basis. Defining
\begin{equation}
    a_\pm=\sqrt{1-p}\cos\theta\pm\sqrt{p}\sin\theta ,
\end{equation}
$K_0^{(\theta,\phi)}$ can be factored, up to a global phase, as
\begin{equation}
    K_0^{(\theta,\phi)}
    = 
    \mathrm{diag}\!\left(1,e^{-i\alpha}\right)\,
    \mathrm{diag}\!\left(A_+,A_-\right),
    \label{eq:K0_phase_decomposition}
\end{equation}
with
\begin{equation}
    \begin{aligned}
    A_\pm &= \sqrt{a_\pm^2\sin^2\phi+a_\mp^2\cos^2\phi}, \\
    \alpha &= \arctan\!\left(\frac{a_+}{a_-}\tan\phi\right)-\arctan\!\left(\frac{a_-}{a_+}\tan\phi\right).
    \end{aligned}
    \label{eq:K0_phase_parameters}
\end{equation}
An analogous form holds for $K_1^{(\theta,\phi)}$. The second factor in Eq.~\eqref{eq:K0_phase_decomposition} is a positive diagonal operator, i.e., a weak measurement of $Z$, while the first factor is a $Z$ rotation by the angle $\alpha$, which is a non-Clifford operation unless $\alpha$ is a multiple of $\pi/2$. For small $\phi$, expanding to first order gives, up to a global phase,
\begin{equation}
    K_0^{(\theta,\phi)}=
    \mathrm{diag}\!\left(1,e^{-i\alpha}\right)\,
    K_0^{(\theta,0)}+\mathcal{O}(\phi^2),
\end{equation}
with $\alpha\simeq{(a_+^2-a_-^2)}/{(a_+a_-)}\,\phi$.
Hence, the Kraus operator acts as its $\phi=0$ counterpart followed by a small non-Clifford $Z$ rotation, which injects magic into the state. At $\phi=\pi/4$, where $M_2$ is numerically observed to be maximal, one instead finds $A_+=A_-$, so that
\begin{equation}
    K_0^{(\theta,\pi/4)}\propto
    \mathrm{diag}\!\left(1,e^{-i\alpha}\right)
\end{equation}
is a pure phase gate. In this case, the unraveling is unitary: it does not condition the state on any measurement outcome, but simply dresses the circuit with additional random non-Clifford rotations, which leave the magic at the Haar value. These limits show that choosing $\phi>0$ generally injects magic, making $\phi=0$ the most relevant case. The bottom row of Fig.~\ref{fig4} therefore shows the cuts at $\phi=0$, where the Kraus operators describe a two-outcome weak measurement of $Z$. On the same plots, we mark the angle $\theta^\ast$ of Eq.~\eqref{eq:thetastar} and the magic obtained with the projective unraveling of Eq.~\eqref{eq:projdeph}.

Three observations follow. First, the magic at $\theta^\ast$ coincides
with that of the projective unraveling within the statistical error,
for all values of $p$. This is the near coincidence anticipated in
Sec.~\ref{sec:noise} and explained in Appendix~\ref{app:thetastar}:
for a single state the two unravelings yield SREs that differ at first
order in $p$, but the difference is odd under a bit flip on the noisy
qubit, and it therefore averages to zero over the bit-flip-symmetric
ensemble of states generated by the random circuit. Second, the
minimum of the magic is attained at an angle $\theta_{\min}>\theta^\ast$, at
which neither Kraus operator is a projector: both branches are weak
measurements of $Z$, of different strengths. Third, and most
importantly, the minimal magic obtained within the unitary family is
\emph{smaller} than that obtained with the projective unraveling. This
result runs against the intuition that projective measurements, which
map the measured qubit onto a stabilizer state and remove its
contribution to the magic, should be the most effective way of
unraveling noise if the goal is to minimize non-stabilizerness, and
against the widespread use of projective measurements as the paradigm
of magic-destroying operations in the literature on monitored
circuits~\cite{bejan2024dynamicalmagictransitions,fux2024entanglementmagicseparation,tirrito2025magicphasetransitionsmonitored,russomanno2025nonstabilizernessintheunitary}. A similar conclusion was reached for entanglement in Ref.~\cite{zihan2023efficientsaqmpling}, where weak measurements were found to be the most disentangling unraveling of unital channels. It also shows that there is room for a non-trivial optimization of the unraveling even in the state-agnostic setting considered here, where the same Kraus operators are used at every noise event regardless of the current state of the trajectory.

\begin{figure}[t]
  \centering
  \includegraphics[width=0.45\textwidth]{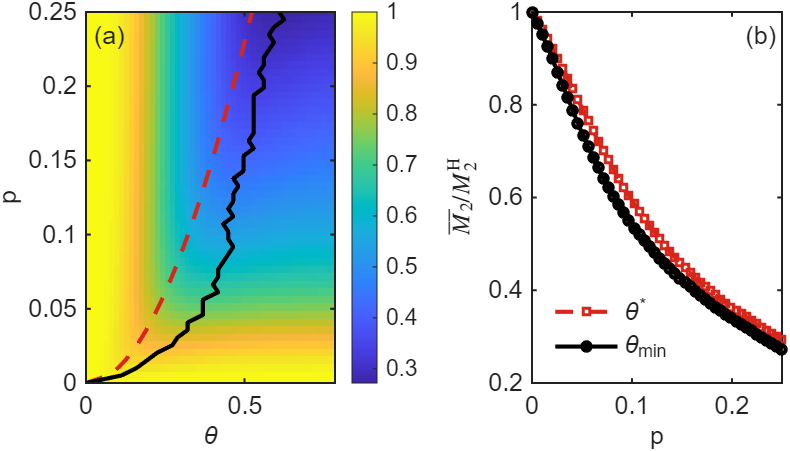}
  \caption{\textbf{Unraveling dependence of magic in noisy Haar-random brickwork circuits.}
(a) $\Mavg/\MH$ as a function of the unraveling angle $\theta$ and the dephasing probability $p$, with $\phi=0$, at the end of circuits with $n=8$ qubits and $N_L=16$ layers, sampling $50$ values each of $\theta$ and $p$. The red dashed line marks the quantum-jump unraveling; the black solid line marks $\theta_{\min}(p)$, the angle minimizing $\Mavg$ among the sampled $\theta$ values at each $p$. (b) $\Mavg/\MH$ at $\theta=\theta^\ast$ (red squares, dashed line) and $\theta=\theta_{\min}$ (black circles, solid line) as a function of $p$. Values at $\theta^\ast$ are obtained by linear interpolation along the sampled $\theta$ grid. All averages use $8$ random circuit realizations, with $1000$ trajectories per realization.}
  \label{fig5}
\end{figure}

Figure~\ref{fig5} summarizes the dependence on $p$. Figure~\ref{fig5}(a)
shows $\Mavg/\MH$ as a function of $\theta$ and $p$ at $\phi=0$, with
the curves $\theta^\ast(p)$ and $\theta_{\min}(p)$ superimposed. The
optimal angle follows the same qualitative dependence on $p$ as
$\theta^\ast$, growing from zero as $\sqrt{p}$ at small $p$, but it
remains systematically larger. Figure~\ref{fig5}(b) shows
the minimal magic $\Mavg(\theta_{\min})$ and the magic at
$\theta^\ast$ as a function of $p$. Both decrease monotonically with
$p$, approximately linearly at small $p$, without any sign of
saturation in the range explored.

\begin{figure*}[t]
  \centering
  \includegraphics[width=0.9\textwidth]{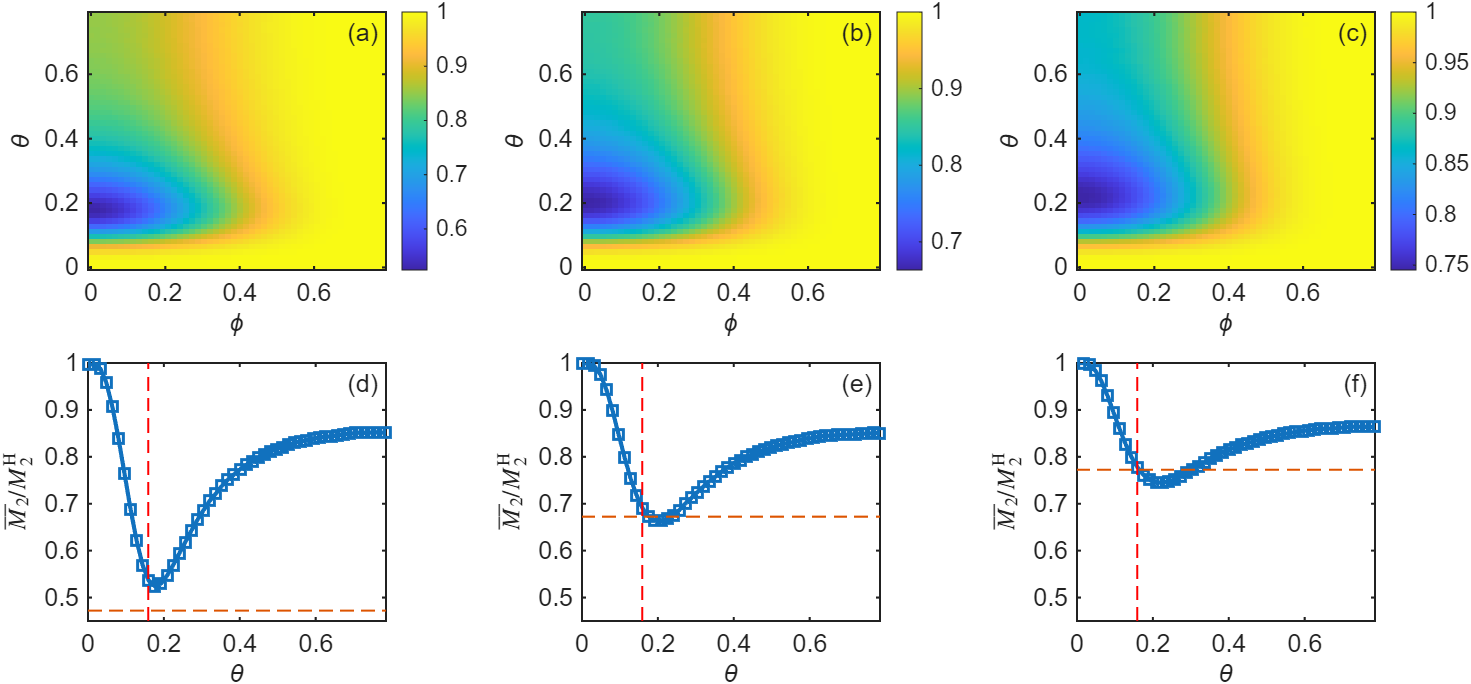}
  \caption{\textbf{Magic in noisy $T$-doped random Clifford brickwork circuits as a function of the unraveling parameters.}
(a)--(c) $\Mavg/\MH$ at the end of the circuit, for dephasing probability $p=0.025$. The $T$ gate probabilities and circuit depths are $q=0.05$, $N_L=48$ in (a),(d); $q=0.2$, $N_L=32$ in (b),(e); and $q=0.6$, $N_L=32$ in (c),(f). Panels (a)--(c) sample $50$ values of each unraveling angle, $\theta$ and $\phi$; panels (d)--(f) show the corresponding $\phi=0$ cuts (symbols connected by solid lines). Vertical red dashed lines mark the quantum-jump unraveling ($\theta=\theta^\ast$). Horizontal dashed lines show the values obtained with the projective unraveling of Eq.~\eqref{eq:projdeph}. The angle scans average over $8$ random circuit realizations, whereas the projective reference values average over $80$ realizations, with $1000$ trajectories per realization in both cases.}
  \label{fig6}
\end{figure*}

\begin{figure}[t]
  \centering
  \includegraphics[width=0.45\textwidth]{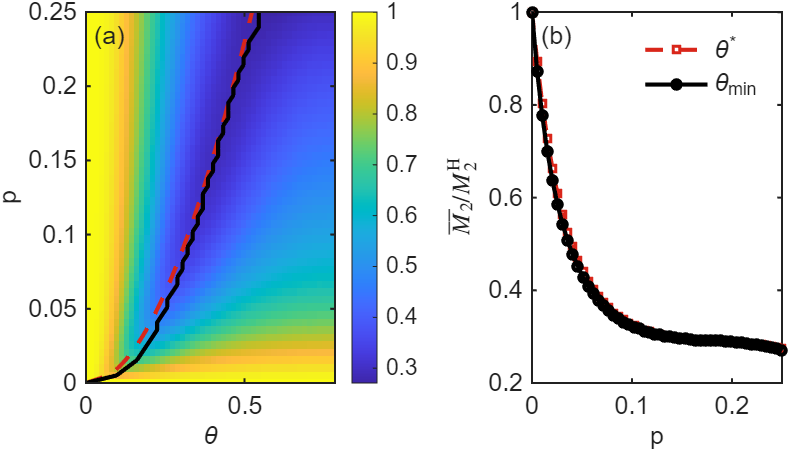}
\caption{\textbf{Unraveling dependence of magic in noisy $T$-doped random Clifford brickwork circuits.}
(a) $\Mavg/\MH$ as a function of the unraveling angle $\theta$ and the dephasing probability $p$, with $\phi=0$, at the end of circuits with $n=8$ qubits, $N_L=32$ layers, and $T$ gate probability $q=0.1$, sampling $50$ values each of $\theta$ and $p$. The red dashed line marks the quantum-jump unraveling; the black solid line marks $\theta_{\min}(p)$, the angle minimizing $\Mavg$ among the sampled $\theta$ values at each $p$. (b) $\Mavg/\MH$ at $\theta=\theta^\ast$ (red squares, dashed line) and $\theta=\theta_{\min}$ (black circles, solid line) as a function of $p$. Values at $\theta^\ast$ are obtained by linear interpolation along the sampled $\theta$ grid. All averages use $8$ random circuit realizations, with $1000$ trajectories per realization.}
\label{fig5b}
\end{figure}

\subsection{$T$-doped random Clifford circuits}
\label{sec:cliffordT}

Haar-random circuits generate magic rapidly~\cite{turkeshi2025magicspreading}, driving the state to near-maximal magic, and in this sense represent a worst-case scenario. Circuits of practical interest, such as those implementing quantum algorithms on fault-tolerant hardware, are instead naturally expressed in the Clifford$+T$ gate set. Their magic is controlled by the number of $T$ gates and is generally far from maximal~\cite{bravyi2016improvedclassicalsimulation,bravyi2019simulationofquantum,paviglianiti2026truecostfactoringlinking}. Random $T$-doped Clifford circuits~\cite{zhou2020singletgate,haferkamp2023efficientunitarydesigns,leone2021quantumchaosis} provide a minimal model for this situation, in which the rate of magic generation is tuned by the doping probability $q$. They are therefore well suited to assess whether the conclusions drawn for Haar-random circuits carry over to circuits of more direct relevance for the simulability of quantum algorithms.

Figure~\ref{fig6} shows the trajectory-averaged magic at the end of noisy $T$-doped Clifford circuits as a function of $\theta$ and $\phi$, for increasing doping $q$ at $p=0.025$. As in the Haar case, the magic is minimal at $\phi=0$ and, along this line, reaches its lowest value at an angle $\theta_{\min}$ slightly larger than $\theta^\ast$, owing to the structure of the Kraus operators discussed in Sec.~\ref{sec:haar}. The reduction of magic at $\theta_{\min}$ is largest at low doping and decreases as $q$ increases, approaching the Haar behavior, although the $T$-doped ensemble does not converge to the Haar ensemble as $q\to1$.
Whether the projective unraveling performs better or worse than the minimal family depends on the doping. At intermediate and high doping (Figs.~\ref{fig6}(e) and~\ref{fig6}(f)), the projective magic is close to that at $\theta^\ast$ and slightly above the minimum of the family, as for Haar-random circuits. At low doping (Fig.~\ref{fig6}(d)), it is lower than that of any unraveling of the family. This can be understood from the Clifford limit $q=0$, in which projective measurements keep every trajectory a stabilizer state with zero magic, whereas the non-Clifford no-click operator $K_0^\ast$ of the $\theta^\ast$ unraveling generates magic (Appendix~\ref{app:thetastar}). By continuity in $q$, this advantage persists at low doping. When magic is scarce, the best unraveling among those considered is therefore not part of the minimal family.

Figure~\ref{fig5b} shows the dependence on $p$ of the optimal unraveling and of the minimal magic in $T$-doped Clifford circuits at fixed $q=0.1$. In contrast with the Haar case, the optimal angle $\theta_{\min}(p)$ follows $\theta^\ast(p)$ closely over the whole range of $p$, and the magic at $\theta^\ast$ is practically indistinguishable from the minimal one. The minimal magic decreases rapidly at small $p$, faster than in the Haar case, and then flattens to a plateau for $p\gtrsim0.1$. This behavior can be understood qualitatively as the outcome of a competition between the injection of magic by the $T$ gates, at a rate set by $q$, and its removal by the measurement-like noise events, at a rate set by $p$.

\subsection{Scaling with system size}
\label{sec:scaling}

\begin{figure}[ht]
  \centering
  \includegraphics[width=\columnwidth]{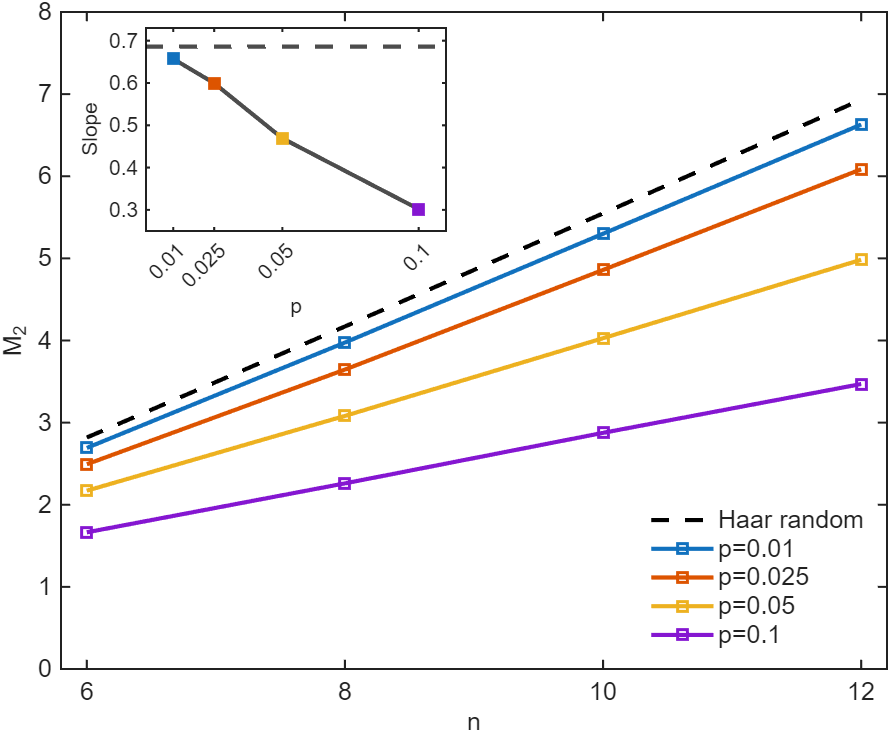}
  \caption{\textbf{Scaling with system size.}
$\Mavg$ at the end of noisy Haar-random circuits with $N_L=2n$ layers, as a function of the number of qubits $n$, for the optimal unraveling $\theta=\theta_{\min}$ at $\phi=0$ and several dephasing strengths $p$. For each $n$ and $p$, $\Mavg(\theta_{\min})$ is taken as the lowest value obtained in a scan of $\theta$ over the interval $[\theta^\ast,2\theta^\ast]$, using $20$ equally spaced values for $n=6$ and $8$, and $10$ values for $n=10$ and $12$. Each value of $\theta$ is an independent calculation, averaging over $36$ random circuit realizations with $1000$ trajectories each for $n=6$, $8$, and $10$, and over $8$ realizations with $200$ trajectories each for $n=12$. The dashed line shows the Haar-random value $\MH$ of Eq.~\eqref{eq:haar_magic}, which is also the value obtained with the canonical unraveling $\theta=0$. Inset: slope of $\Mavg$ versus $n$, obtained from a linear fit, as a function of $p$; the dashed line marks the asymptotic Haar-random slope $\ln 2$ per qubit.}
  \label{fig7}
\end{figure}

Having established that a suitable unraveling reduces the average
magic of the trajectories, we now ask how this reduction scales with
the system size. Fig.~\ref{fig7} shows $\Mavg$ at the end of noisy
Haar-random circuits for the optimal unraveling, as a function of $n$
between $6$ and $12$ and for several values of $p$, together with the
Haar-random value $\MH$, which coincides with the result of the
canonical unraveling. For all $p$, the minimal magic grows linearly
with $n$ within our numerical accuracy: the optimal unraveling does not
change the extensive character of the non-stabilizerness of the
trajectories. 

While the accessible system sizes are too small to exclude subleading
corrections, the data indicate that the unraveling optimization
achieves a constant-factor reduction of the magic density, and not a
change of its scaling. A typical trajectory of the optimally unraveled dynamics thus still carries an amount of magic proportional to the system size, so that its simulation by stabilizer-expansion methods remains exponentially costly in $n$. Whether the smaller magic density also lowers the exponential growth rate of the average cost depends on the full distribution of magic over trajectories, which we analyze next.

\subsection{Distribution of magic over trajectories}
\label{sec:distribution}

The average magic is only part of the story. The computational cost of a trajectory-based classical simulation is determined by the distribution of the resource over the ensemble of trajectories, and, when the cost per trajectory grows exponentially with the resource, by its tail. Recent studies of entanglement in different unravelings have shown that, while the average entanglement can be made small by an optimal choice of unraveling~\cite{vovk2022entanglementoptimaltrajectories,vovk2024quantumtrajectoryentanglement,chen2024optimizedtrajectoryunraveling,daraban2025nonunitaritymaximizing}, the fluctuations across trajectories can become large, so that the actual computational advantage results from a trade-off between the average cost per trajectory and the number of trajectories needed for a given statistical accuracy~\cite{cichy2026classicalsimulationofnoisy,sander2026computationalregimesmatrixproductstatebasedquantum}. Here we show that the optimal unraveling for magic likewise produces a broad distribution of magic over trajectories.

\begin{figure*}[t!]
  \centering
  \includegraphics[width=0.9\textwidth]{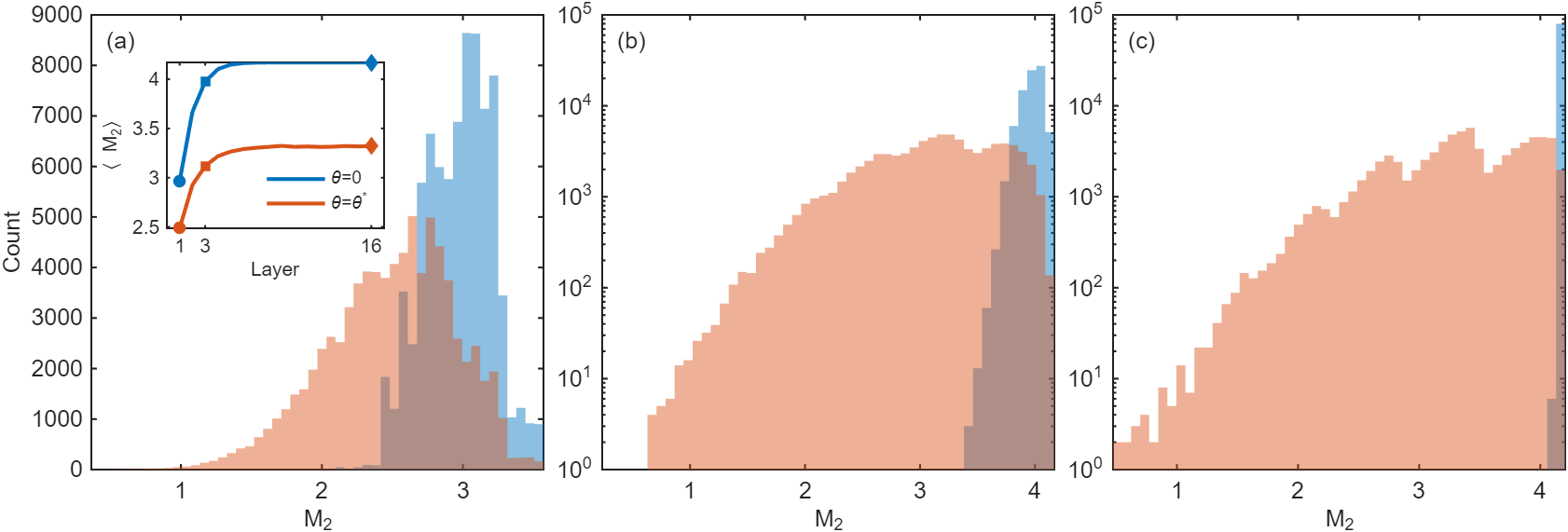}
\caption{\textbf{Distribution of magic along noisy Haar-random brickwork circuits.}
Histograms of $M_2$ after (a) $1$, (b) $3$, and (c) $16$ layers, for the canonical unraveling ($\theta=0$, blue) and the quantum-jump unraveling ($\theta=\theta^\ast$, orange), with $\phi=0$ in both cases. Circuits contain $n=8$ qubits and $N_L=16$ layers, with dephasing probability $p=0.05$. Each histogram pools $1000$ trajectories for each of $80$ random circuit realizations, giving $80\,000$ samples, using $50$ equally spaced bins with shared edges for the two unravelings within each panel. The vertical axis shows counts on a linear scale in (a) and a logarithmic scale in (b),(c). Inset in (a): $\Mavg$, not normalized, after each layer. Circles, squares, and diamonds mark the layers shown in (a), (b), and (c), respectively; colors identify the same unravelings as in the histograms.}
\label{fig8}
\end{figure*}

\begin{figure}[t!]
  \centering
  \includegraphics[width=\columnwidth]{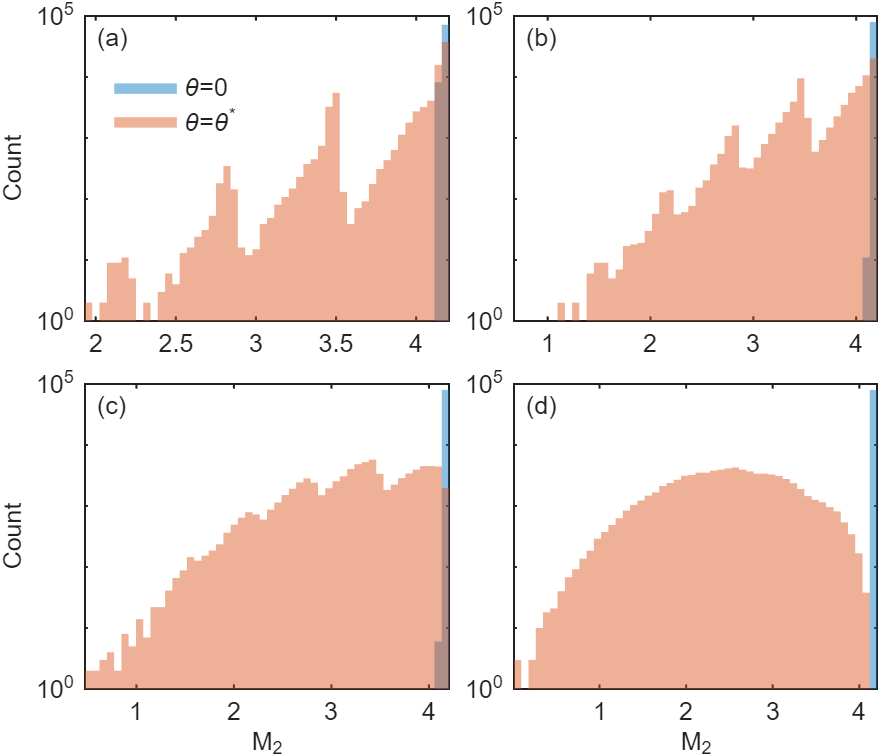}
  \caption{\textbf{Distribution of magic at the end of noisy Haar-random brickwork circuits.}
Histograms of $M_2$ after $N_L=16$ layers for $n=8$ qubits, with dephasing probabilities (a) $p=0.01$, (b) $p=0.025$, (c) $p=0.05$, and (d) $p=0.1$. Each panel compares the canonical unraveling ($\theta=0$, blue) with the quantum-jump unraveling ($\theta=\theta^\ast$, orange), with $\phi=0$ in both cases. Each histogram pools $1000$ trajectories for each of $80$ random circuit realizations, giving $80\,000$ samples, using $50$ equally spaced bins with shared edges for the two unravelings within each panel. The vertical axes show counts on a logarithmic scale.}
  \label{fig9}
\end{figure}

Figure~\ref{fig8} shows how the distribution of $M_2$ over trajectories evolves along a noisy Haar-random circuit, for the canonical unraveling and for the quantum-jump unraveling at $\theta^\ast$. For the canonical unraveling, the distribution is narrow at all times and, at saturation, sharply peaked at the Haar value $\MH$. This reflects the concentration of the stabilizer R\'enyi entropy of random states around its typical value~\cite{leone2022stabilizerrenyientropy,turkeshi2025paulispectrum,szombathy2025asymptoticallyindependent}, which the Clifford noise events cannot spoil. For the $\theta^\ast$ unraveling, instead, the distribution is already broader after the first layer and becomes extremely broad at saturation. It extends from a small fraction of $\MH$ up to values close to $\MH$, with most of the weight concentrated in the upper part of this range. In other words, although the average magic is reduced, a sizable fraction of the trajectories carries a magic close to the full Haar value, while the reduction of the average comes from a long tail of trajectories with much lower magic. Note that the distributions in Figs.~\ref{fig8}
and \ref{fig9} are accumulated over $80$ circuit realizations; we
verified that the broadening is a property of the trajectory ensemble
of each individual circuit, and not a consequence of
circuit-to-circuit fluctuations.

Figure~\ref{fig9} shows the distribution at the end of the circuit for increasing $p$. The distribution remains broad over the whole range of noise strengths. At small $p$, it is dominated by a peak at the Haar value, which collects the trajectories in which no click occurred in the last layers, followed by a comb of peaks separated by approximately $\ln 2$. These correspond to trajectories in which $1,2,3,\dots$ clicks occurred in the last layer of the circuit, each click mapping one qubit onto a $Z$ eigenstate and thereby removing about one qubit's worth of Haar-random magic. As $p$ increases, the number of clicks per layer grows, and the peaks merge into a smooth, broad distribution centered well below the Haar value, with a tail extending up to it. Correspondingly, the weight of the near-Haar peak decreases markedly with $p$. For all $p$, the distribution is much broader than the canonical one, and its width is comparable to the reduction of the average magic. The peak structure also identifies the origin of the high-magic tail: its weight is set by the probability that no click occurs in the last layers of the circuit.

These results have direct consequences for classical simulation. For a stabilizer-expansion method, the cost of processing a state grows exponentially with its magic~\cite{bravyi2016improvedclassicalsimulation,bravyi2019simulationofquantum,seddon2021quantifyingquantumspeedups,bittel2026operationalinterpretation}. The average cost over the trajectory ensemble is then dominated by the high-magic tail of the distribution rather than by its mean. Trajectories with low magic, which are cheap to simulate, contribute little to the average of an exponential cost, whereas trajectories that retain near-Haar magic are as costly as in the canonical unraveling. The reduction of the average magic reported in the previous sections therefore does not translate into a comparable reduction of the simulation cost. The optimal unraveling still offers an advantage, however, because these costly trajectories are sampled less often than in the canonical unraveling, where every trajectory carries near-Haar magic. If a fraction $w$ of trajectories retains near-Haar magic, the cost is reduced by at most a factor $1/w$. Since the weight of the near-Haar peak in Fig.~\ref{fig9} decreases markedly with $p$, the attainable gain grows with the noise strength. Whether it persists for larger systems depends on how this weight scales with $n$, which we leave for future work. Unless the high-magic trajectories turn out to be irrelevant for the observables of interest, this bound sets the attainable advantage.

\section{Discussion and Conclusions}
\label{sec:discussion}

In this work, we have investigated how the magic of quantum trajectories depends on the unraveling of the noise channel in noisy random quantum circuits. Our results show that this magic is not fixed by the noisy dynamics, but can be substantially reduced through an appropriate choice of unraveling, even in a state-agnostic way. The optimal unraveling differs between Haar-random and weakly doped Clifford circuits, but in both cases magic remains extensive. The achieved reduction of the magic density depends on the noise strength, but the mean alone does not capture the full picture of the complexity of typical realizations: the distribution of magic over trajectories becomes broad, and at weak noise a sizable fraction of trajectories retains near-maximal magic. Since the cost of stabilizer-based simulation grows exponentially with magic, these trajectories control the simulation cost, which remains exponential in the system size. They are, however, sampled less often than in the canonical unraveling, which indicates a potential gain in efficiency, whose size depends on how their weight scales with the system size.

Our results can be read from two complementary points of view: that of
resource theory and that of classical simulation.

From the resource-theoretic point of view, the unraveling of a channel is a choice of pure-state decomposition of the density matrix, and the trajectory-averaged magic is a property of that decomposition rather than of the state. This is exactly the structure underlying the convex-roof extension of a pure-state resource measure to mixed states~\cite{chitambar2019quantumresourcetheories,leone2024stabilizerentropiesaremonotones}. The convex roof of $M_2$, evaluated on the density matrix at the end of the circuit, is the minimum of the average magic over \emph{all} pure-state decompositions of that density matrix. It therefore lower-bounds the trajectory-averaged magic of any unraveling, including adaptive ones that select the Kraus operators as a function of the current state. The state-agnostic optimization performed here explores only a restricted set of decompositions, namely those generated by a fixed local Kraus representation applied at every noise event. Since the minimum over this set is already well below the Haar value, a substantial part of the magic of the canonical trajectories must be due to the choice of decomposition, and not to the mixed state itself.

However, this minimum remains extensive. It is an open question whether the residual magic is intrinsic to the noisy state, i.e., whether the convex roof itself is extensive, or whether it could be further reduced by more general decompositions, such as adaptive or nonlocal unravelings. The former scenario would be consistent with the expectation that local dephasing at a finite rate per gate can only remove a finite fraction of the magic generated by an extensive number of gates per layer, and with recent evidence that the magic of states generated by random circuits is highly robust to depolarizing noise~\cite{haug2026efficientwitnessing}. Estimating how far the optimal local unraveling is from the convex roof would settle this point and quantify the room left for improved unravelings.

The structure of the optimal unraveling is also instructive. The canonical unraveling in terms of Pauli jumps is the natural choice for a stabilizer simulator, because its noise events are free operations, but it is the least useful as far as the magic of the trajectories is concerned: being Clifford, it cannot remove any magic. At the opposite extreme, the projective unraveling maps the noisy qubit onto a $Z$ eigenstate at every click, and projective measurements are the paradigmatic magic-reducing operation in the literature on monitored circuits~\cite{bejan2024dynamicalmagictransitions,fux2024entanglementmagicseparation,tirrito2025magicphasetransitionsmonitored}. The optimal unraveling we find is in between, since both of its Kraus operators are weak measurements, and, in the high-magic regime, it outperforms the projective unraveling. This parallels the finding of Ref.~\cite{zihan2023efficientsaqmpling} that weak measurements are the most disentangling unraveling of unital noise channels, and suggests that, when the resource is abundant, weak measurements are more effective than projective ones at reducing nonlinear resources, whether entanglement or magic. In the low-magic regime of weakly doped Clifford circuits, however, the projective unraveling gives the lowest magic among the unravelings considered, so the best unraveling depends on how much magic the state carries. This is a first indication that adaptive strategies could outperform state-agnostic ones.

From the point of view of classical simulation, the picture is more nuanced than the reduction of the average magic suggests. In Haar-random circuits, the optimal unraveling lowers the magic density of the trajectories by a factor that depends on the noise strength, while in $T$-doped Clifford circuits it brings the average magic down to a $q$-dependent plateau that, for $p\gtrsim0.1$, depends only weakly on the noise strength. Since the average magic remains extensive, a stabilizer-based simulation remains exponentially costly, and the optimal unraveling cannot make it efficient. Moreover, the broad distribution of magic over trajectories implies that the mean is not the relevant quantity: the simulation cost is controlled by the high-magic trajectories, which at weak noise retain near-Haar magic with sizable probability. These trajectories are, however, sampled less often than in the canonical unraveling, so the attainable gain is set by their weight and by how it scales with the system size. A similar trade-off has recently been found for unravelings that minimize entanglement~\cite{cichy2026classicalsimulationofnoisy,sander2026computationalregimesmatrixproductstatebasedquantum}. In both cases, the lesson for trajectory-based classical algorithms is the same: an unraveling that minimizes the average resource does not necessarily minimize the simulation cost.

Several questions remain open. First, it is unclear how the broad distribution of magic affects the estimation of physical observables. As discussed in Sec.~\ref{sec:distribution}, the high-magic trajectories of the quantum-jump unraveling are those in which no click occurred in the last layers. If their contribution to quasi-local observables is close to that of the low-magic trajectories, a truncated or importance-sampled average over trajectories could substantially reduce the simulation cost, at the price of a controlled bias.

Second, our optimization is state-agnostic. Adaptive unravelings, which choose the Kraus representation at each noise event depending on the current state, have proven very effective at reducing entanglement~\cite{vovk2022entanglementoptimaltrajectories,vovk2024quantumtrajectoryentanglement,chen2024optimizedtrajectoryunraveling,daraban2025nonunitaritymaximizing,cichy2026classicalsimulationofnoisy}. For magic, an adaptive strategy could specifically target the high-magic trajectories, and thus reduce not only the average magic but also the weight of the high-magic tail, which ultimately sets the simulation cost. A natural candidate is to choose, at each noise event, the local Kraus representation that minimizes the expected value of $M_2$ after that step. The cost of evaluating $M_2$ for each candidate could be reduced with the stochastic estimators mentioned in Sec.~\ref{sec:magic}.

Third, while our main analysis focuses on dephasing noise, the results of Appendix~\ref{app:noise} show that the same behavior holds for depolarizing noise, at least within the restricted family of unravelings considered there, and suggest that it should extend to any channel whose canonical Kraus operators are Clifford. Amplitude damping, on the other hand, cannot be unraveled into Clifford unitaries at all, since it is not unital, and within the minimal family we find no reduction of magic with respect to the canonical unraveling. Whether more general unravelings, such as nonminimal ones, can reduce the magic of its trajectories remains to be seen.

Fourth, the exact evaluation of $M_2$ limits us to $n\leq12$. This is enough to show that the minimal magic grows linearly with $n$, but not to exclude subleading corrections. Stochastic estimators of $M_2$ for MPS would allow larger sizes at strong noise, where the entanglement of the trajectories is also expected to be reduced.

Finally, we have studied magic alone, while many classical simulation methods are limited by both magic and entanglement. The unraveling that minimizes one resource need not minimize the other, so the best choice for a given classical algorithm depends on which resource controls its cost~\cite{fux2024entanglementmagicseparation,paviglianiti2026emergenceofgenericentanglement,fioroni2026noiseinduced}. How the two resources depend jointly on the unraveling, and which unraveling is best for methods sensitive to both, remains to be understood.

\begin{acknowledgments}
We acknowledge support from the Swiss National Science Foundation through Projects No. 200020\_215172, 200021-227992, and 20QU-1\_215928, and as a part of NCCR SPIN (grant number 225153).
\end{acknowledgments}

\bibliography{biblio}

\newpage
\appendix
\onecolumngrid

\section{Depolarizing and amplitude damping noise models}
\label{app:noise}

The construction of Sec.~\ref{sec:noise} extends naturally to other
single-qubit channels. In this appendix we describe the corresponding
families of unravelings for the depolarizing and amplitude-damping
channels, and we present numerical results for the trajectory-averaged
magic of Haar-random and $T$-doped random Clifford circuits subject to
these two channels, obtained with the same protocol as in the main
text. We first consider the single-qubit depolarizing channel,
\begin{equation}
    \mathcal{E}^{\mathrm{dep}}_p(\rho)
    =
    (1-p)\rho
    +
    \frac{p}{3}
    \sum_{\alpha=x,y,z}
    \sigma_\alpha\rho\sigma_\alpha ,
    \label{eq:depolarizing_channel}
\end{equation}
where
$(\sigma_x,\sigma_y,\sigma_z)=(X,Y,Z)$. Its canonical Kraus
representation is
\begin{equation}
    K_0=\sqrt{1-p}\,I,
    \qquad
    K_\alpha=\sqrt{\frac{p}{3}}\,\sigma_\alpha,
    \quad \alpha=x,y,z .
    \label{eq:canonical_depolarizing}
\end{equation}
To construct a simple two-parameter family of unravelings, we first
split the identity contribution into three identical Kraus operators.
For each $\alpha\in\{x,y,z\}$, define
\begin{equation}
    K_{\alpha,0}
    =
    \sqrt{\frac{1-p}{3}}\,I,
    \qquad
    K_{\alpha,1}
    =
    \sqrt{\frac{p}{3}}\,\sigma_\alpha .
    \label{eq:split_depolarizing}
\end{equation}
The six operators in Eq.~\eqref{eq:split_depolarizing} still represent
the channel in Eq.~\eqref{eq:depolarizing_channel}. We then apply the
same unitary transformation to each of the three pairs:
\begin{align}
    K_{\alpha,0}^{(\theta,\phi)}
    &=
    e^{i\phi}\sqrt{\frac{1-p}{3}}\cos\theta\,I
    -
    e^{-i\phi}\sqrt{\frac{p}{3}}\sin\theta\,\sigma_\alpha,
    \nonumber\\
    K_{\alpha,1}^{(\theta,\phi)}
    &=
    e^{i\phi}\sqrt{\frac{1-p}{3}}\sin\theta\,I
    +
    e^{-i\phi}\sqrt{\frac{p}{3}}\cos\theta\,\sigma_\alpha .
    \label{eq:depolarizing_unraveling_explicit}
\end{align}
This construction is deliberately nonexhaustive: a general unraveling
could use different rotations for the three Pauli sectors or mix Kraus
operators belonging to different sectors.

A projective-measurement unraveling can also be constructed for the
depolarizing channel. Let
\begin{equation}
    P_{\alpha,\pm}
    =
    \frac{I\pm\sigma_\alpha}{2}
\end{equation}
be the projectors onto the two eigenstates of $\sigma_\alpha$. By applying the same rectangular isometry to the three pairs of Kraus operators \eqref{eq:split_depolarizing}, one obtains the seven Kraus operators
\begin{equation}
    K_{\mathrm{id}}=\sqrt{1-2p}\,I,
    \qquad
    K_{\alpha,\pm}
    =
    \sqrt{\frac{2p}{3}}\,P_{\alpha,\pm},
    \quad \alpha=x,y,z ,
    \label{eq:projdep}
\end{equation}
and the channel is written as
\begin{equation}
    \mathcal{E}^{\mathrm{dep}}_p(\rho)
    =
    (1-2p)\rho
    +
    \frac{2p}{3}
    \sum_{\alpha=x,y,z}
    \sum_{s=\pm}
    P_{\alpha,s}\rho P_{\alpha,s}.
    \label{eq:depolarizing_measurement_decomposition}
\end{equation}
The nonnegative coefficient of the identity outcome requires
$0\leq p\leq1/2$. At $p=1/2$, that outcome has zero probability and
the unraveling consists solely of the six Pauli-measurement outcomes.

\begin{figure}[ht]
  \centering
  \includegraphics[width=0.45\textwidth]{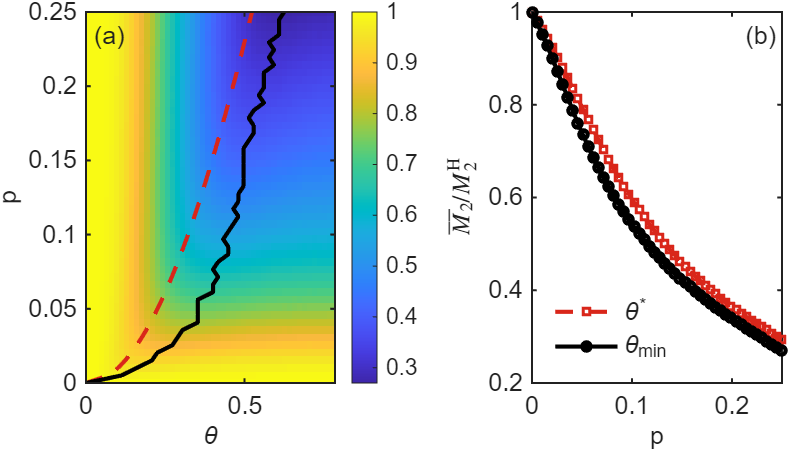}
  \caption{\textbf{Optimal unraveling in noisy Haar-random circuits.}
Normalized trajectory-averaged stabilizer R\'enyi entropy
$\Mavg/\MH$ at the end of Haar-random brickwork circuits with
depolarizing noise, for $n=8$ qubits and $N_L=16$ layers.
Results are averaged over $8$ independent circuit realizations
and $1000$ quantum trajectories per realization.
(a) $\Mavg/\MH$ as a function of the unraveling angle $\theta$
and the noise strength $p$, at fixed $\phi=0$, sampled on a
$50\times50$ grid.
The dashed red line denotes the quantum-jump unraveling,
$\theta^\ast(p)$, and the solid black line
denotes $\theta_{\min}(p)$, the angle minimizing the averaged
entropy over the sampled $\theta$ values at each $p$.
(b) $\Mavg/\MH$ evaluated at $\theta^\ast(p)$
(red squares, dashed line) and $\theta_{\min}(p)$
(black circles, solid line), as a function of $p$.
Values at $\theta^\ast(p)$ are obtained by linear interpolation
along the $\theta$ direction at fixed $p$.}
  \label{fig10}
\end{figure}

\begin{figure}[ht]
  \centering
  \includegraphics[width=0.45\textwidth]{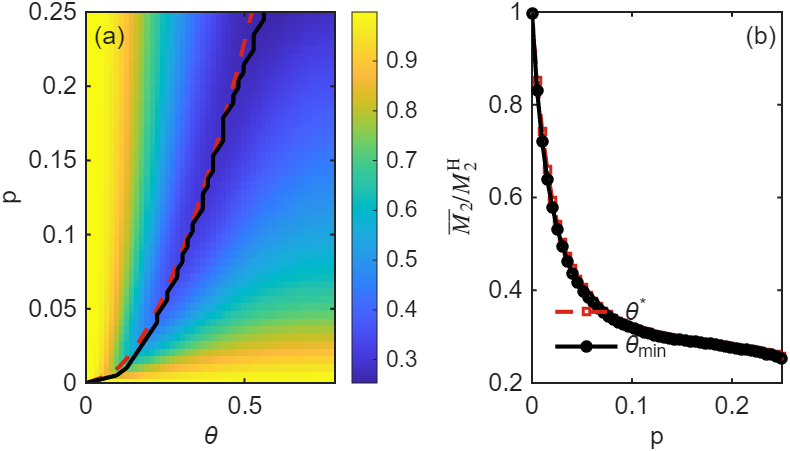}
\caption{\textbf{Optimal unraveling in noisy $T$-doped random Clifford circuits.}
Normalized trajectory-averaged stabilizer R\'enyi entropy
$\Mavg/\MH$ at the end of $T$-doped random Clifford brickwork
circuits with depolarizing noise, for $n=8$ qubits,
$N_L=48$ layers, and $q=0.05$.
Results are averaged over $8$ independent circuit realizations
and $1000$ quantum trajectories per realization.
(a) $\Mavg/\MH$ as a function of the unraveling angle $\theta$
and the noise strength $p$, at fixed $\phi=0$, sampled on a
$50\times50$ grid.
The dashed red line denotes
$\theta^\ast(p)$, and the solid black line
denotes $\theta_{\min}(p)$, the angle minimizing the averaged
entropy over the sampled $\theta$ values at each $p$.
(b) $\Mavg/\MH$ evaluated at $\theta^\ast(p)$
(red squares, dashed line) and $\theta_{\min}(p)$
(black circles, solid line), as a function of $p$.
Values at $\theta^\ast(p)$ are obtained by linear interpolation
along the $\theta$ direction at fixed $p$.}
  \label{fig11}
\end{figure}

Figures~\ref{fig10} and~\ref{fig11} show the trajectory-averaged
magic at the end of noisy Haar-random and $T$-doped random Clifford
circuits, respectively, as a function of the unraveling angle $\theta$
and of the noise strength $p$, for the family of
Eq.~\eqref{eq:depolarizing_unraveling_explicit} at $\phi=0$, together
with the magic at the quantum-jump angle $\theta^\ast(p)$ and at the
numerically determined optimal angle $\theta_{\min}(p)$. We have also
evaluated the projective unraveling of Eq.~\eqref{eq:projdep}. The
results are in all respects similar to those obtained for dephasing
noise in Sec.~\ref{sec:results}: the canonical unraveling, being
Clifford, leaves the magic at its Haar-random value, and all the
observations made for dephasing, concerning the structure of the
optimal unraveling, its relation to the quantum-jump angle
$\theta^\ast$, the comparison with the projective unraveling, and the
dependence on the noise strength, carry over to the depolarizing
channel. The only quantitative difference is that depolarizing noise
yields an even lower trajectory-averaged magic than dephasing noise at
the same value of $p$. We stress, however, that the family of
Eq.~\eqref{eq:depolarizing_unraveling_explicit}, in which the same
$2\times2$ rotation is applied to each Pauli sector, is not general,
and that a more general unitary or isometric unraveling of the
depolarizing channel may lead to different results. Its exploration is
beyond the scope of this work.

\begin{figure}[ht]
  \centering
  \includegraphics[width=0.45\textwidth]{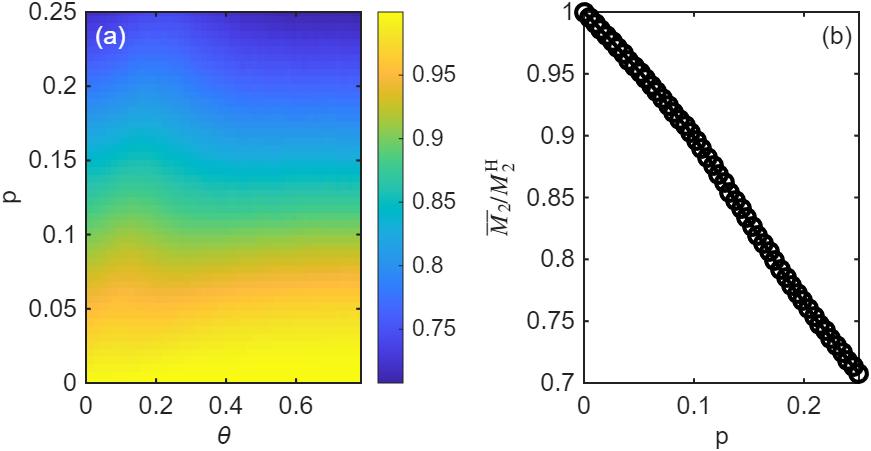}
\caption{\textbf{Optimal unraveling in Haar-random circuits with amplitude damping.}
Normalized trajectory-averaged stabilizer R\'enyi entropy
$\Mavg/\MH$ at the end of Haar-random brickwork circuits with
amplitude-damping noise, for $n=8$ qubits and $N_L=16$ layers.
Results are averaged over $8$ independent circuit realizations
and $1000$ quantum trajectories per realization.
(a) $\Mavg/\MH$ as a function of the unraveling angle $\theta$
and the damping strength $p$, at fixed $\phi=0$, sampled on a
$50\times50$ grid.
(b) Minimum $\Mavg/\MH$ over the sampled
$\theta$ values at each $p$, shown as black circles connected
by a solid line. The minimization is performed at fixed $\phi=0$.}
  \label{fig12}
\end{figure}

\begin{figure}[ht]
  \centering
  \includegraphics[width=0.45\textwidth]{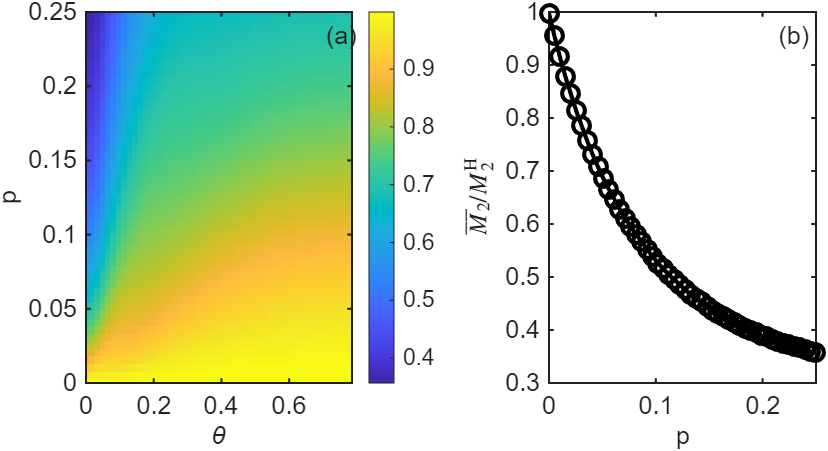}
\caption{\textbf{Optimal unraveling in $T$-doped random Clifford circuits with amplitude damping.}
Normalized trajectory-averaged stabilizer R\'enyi entropy
$\Mavg/\MH$ at the end of $T$-doped random Clifford brickwork
circuits with amplitude-damping noise, for $n=8$ qubits,
$N_L=48$ layers, and $q=0.05$.
Results are averaged over $8$ independent circuit realizations
and $1000$ quantum trajectories per realization.
(a) $\Mavg/\MH$ as a function of the unraveling angle $\theta$
and the damping strength $p$, at fixed $\phi=0$, sampled on a
$50\times50$ grid.
(b) Minimum $\Mavg/\MH$ over the sampled
$\theta$ values at each $p$, shown as black circles connected
by a solid line. The minimization is performed at fixed $\phi=0$.}
  \label{fig13}
\end{figure}

Finally, we consider amplitude-damping noise,
\begin{equation}
    \mathcal{A}_p(\rho)
    =
    A_0\rho A_0^\dagger
    +
    A_1\rho A_1^\dagger,
    \label{eq:amplitude_damping_channel}
\end{equation}
with canonical Kraus operators
\begin{equation}
    A_0
    =
    \begin{pmatrix}
        1 & 0\\
        0 & \sqrt{1-p}
    \end{pmatrix},
    \qquad
    A_1
    =
    \begin{pmatrix}
        0 & \sqrt{p}\\
        0 & 0
    \end{pmatrix}
    =
    \sqrt{p}\,\ket{0}\!\bra{1}.
    \label{eq:amplitude_damping_kraus}
\end{equation}
Here $p$ is the probability for the decay
$\ket{1}\rightarrow\ket{0}$. As for dephasing, all minimal
two-outcome unravelings of this channel can be obtained, up to
outcome-dependent phases, from
\begin{equation}
    \begin{pmatrix}
        A_0^{(\theta,\phi)}\\
        A_1^{(\theta,\phi)}
    \end{pmatrix}
    =
    U(\theta,\phi)
    \begin{pmatrix}
        A_0\\
        A_1
    \end{pmatrix}.
    \label{eq:amplitude_damping_unraveling}
\end{equation}
Note that, at variance with the dephasing and depolarizing channels,
the amplitude-damping channel has no unraveling in terms of Clifford
operations: $A_0$ is a non-unitary, non-Clifford operator for any
$0<p<1$, and $A_1$ is proportional to a non-Hermitian jump operator.
The canonical unraveling therefore already modifies the magic of the
trajectories, and the role of the unraveling optimization can be
expected to differ from the dephasing case. This is confirmed by
Figs.~\ref{fig12} and~\ref{fig13}, which show the trajectory-averaged
magic at the end of noisy Haar-random and $T$-doped random Clifford
circuits, respectively, across the minimal family of
Eq.~\eqref{eq:amplitude_damping_unraveling} at $\phi=0$; for this
channel we have not considered nonminimal unravelings. In contrast to
the dephasing and depolarizing channels, the average magic depends
only weakly on the unraveling angle $\theta$ over the whole range of
damping strengths. Moreover, in both the Haar-random and the
$T$-doped case, the unraveling yielding the minimal average magic
appears to be the canonical one, $\theta=0$, so that no reduction of
the trajectory-averaged magic is obtained by optimizing the unraveling
of amplitude-damping noise within this family.

\section{Projective vs $\theta^\ast$-rotation unravelings}
\label{app:thetastar}

As shown in Fig.~\ref{fig4}(d)--(f), the SRE of the projective unraveling appears to match the one found for the rotation unraveling with $\theta=\theta^\ast$ and $\phi=0$. Here we show that the two values are indeed equal up to first order in $p$, and their difference is $\mathcal{O}(p^2)$. The following derivation assumes Haar-random gates for simplicity, but the result extends to the $T$-doped Clifford case as well, as argued in the end.

For $\theta = \theta^\ast$ and $\phi = 0$, Eq.~\eqref{eq:dephasing_unraveling_explicit} reduces to Eq.~\eqref{eq:thetastar_kraus}, i.e.,
\begin{align}
    &K_0^\ast =  I - 2p P_+,\nonumber\\
    &K_1^\ast = 2\sqrt{p(1-p)}P_+.
\end{align}
Let us consider a state $\ket{\psi}$ and evolve it with these Kraus operators with their Born probabilities. Without any loss of generality, we assume that the noise acts on the first site. Defining $z_1 = \bra{\psi}Z_1\ket{\psi}$, the average final magic is given by
\begin{equation}
    M^\ast_{2,f} = (1-2p -2pz_1)M_2[\ket{\psi^*_0}] + 2p(1+z_1)M_2[\ket{\psi_+}] + \mathcal{O}(p^2),
\end{equation}
where $\ket{\psi_0^\ast} = \ket{\psi} - p (Z_1-z_1)\ket{\psi} + \mathcal{O}(p^2)$ and $\ket{\psi_+} = (1+Z_1)\ket{\psi}/\sqrt{2(1+z_1)}$ are the normalized states obtained after applying $K_0^\ast$ and $K_1^\ast$, respectively. For $p\ll 1$, $M_2[\ket{\psi^*_0}]$ can be expanded around $M_2[\ket{\psi}]$, yielding
\begin{align}
    M_2[\ket{\psi^*_0}] &= M_2[\ket{\psi}] -4p\left(2 z_1-\frac{\sum_{P\in\mathcal{P}_n}\bra{\psi}P\ket{\psi}^3 \bra{\psi}\{P,Z_1\}\ket{\psi}}{\sum_{P\in\mathcal{P}_n}\bra{\psi}P\ket{\psi}^4}\right)+\mathcal{O}(p^2)\nonumber\\
    &= M_2[\ket{\psi}] -4p\left(2 z_1-\frac{\sum_{P_1\in\{I_1,Z_1\}}\sum_{Q\in\mathcal{P}_{n-1}}\bra{\psi}P_1\otimes Q\ket{\psi}^3 \bra{\psi}(P_1 Z_1)\otimes Q\ket{\psi}}{\sum_{P\in\mathcal{P}_n}\bra{\psi}P\ket{\psi}^4}\right)+\mathcal{O}(p^2),
\label{eq:M2_expansion}
\end{align}
where we used $\{X,Z\} = \{Y,Z\} = 0$. We now consider the average of $M^\ast_{2,f}$ over the ensemble of all possible states $\ket{\psi}$ generated by our protocol. Immediately before a noise layer, our setup applies Haar-random unitary gates. This implies that for any possible state $\ket{\psi}$ reached, the state $\ket{\psi'} = X_1\ket{\psi}$ is equally likely. When comparing how Eq.~\eqref{eq:M2_expansion} changes by using $\ket{\psi'}$ instead of $\ket{\psi}$ as the starting state, we observe that both $z_1$ and $\bra{\psi}P_1\otimes Q\ket{\psi}^3 \bra{\psi}(P_1 Z_1)\otimes Q\ket{\psi}$ gain a minus sign, whereas $\bra{\psi'}P\ket{\psi'}^4 = \bra{\psi}P\ket{\psi}^4$ (and thus $M_2[\ket{\psi'}] = M_2[\ket{\psi}]$ as well). As a consequence, several terms vanish on average, leaving
\begin{equation}
    \mathbb{E}[M^\ast_{2,f}] = (1-2p) \mathbb{E}\left[M_2[\ket{\psi}] \right] +2p\mathbb{E}\left[(1+z_1)M_2[\ket{\psi_+}]\right] +\mathcal{O}(p^2).
\end{equation}

Let us now consider the projective unraveling of Eq.~\eqref{eq:projdeph}. For this, the average final magic is
\begin{equation}
    M^\mathrm{proj}_{2,f} = (1-2p)M_2[\ket{\psi}] + p(1+z_1)M_2[\ket{\psi_+}]+p(1-z_1)M_2[\ket{\psi_-}],
\end{equation}
where $\ket{\psi_-} = (1-Z_1)\ket{\psi}/\sqrt{2(1-z_1)}$ is defined analogously to $\ket{\psi_+}$. The difference between the two unravelings, averaged over the ensemble, reads
\begin{equation}
    \mathbb{E}[\delta M_{2,f}] = \mathbb{E}[M^\ast_{2,f}-M^\mathrm{proj}_{2,f}]= p \mathbb{E}[M_2[\ket{\psi_+}]-M_2[\ket{\psi_-}]] + p \mathbb{E}[z_1(M_2[\ket{\psi_+}]+M_2[\ket{\psi_-}])] + \mathcal{O}(p^2).
\end{equation}
Leveraging again the state $\ket{\psi'}$, it is easy to check that $M_2[\ket{\psi'_\pm}] = M_2[\ket{\psi_{\mp}}]$ (where $\ket{\psi'_\pm}$ are defined analogously to $\ket{\psi_\pm}$). This makes the terms of order $p$ vanish, leading to $\mathbb{E}[\delta M_{2,f}] = \mathcal{O}(p^2)$.

This proves that, starting from a common state ensemble, the two unravelings yield equal magic up to $\mathcal{O}(p^2)$ discrepancies. However, the ensembles generated by the two unravelings are not equal. In principle, this could imply that an $\mathcal{O}(p)$ non-stabilizerness difference might arise at later times, due to the ensembles drifting apart. To show this is not the case, let us assume a starting common ensemble of states $\ket{\psi}$, and consider the magic after a noise layer on site $i$, a unitary layer, and another noise layer on site $j$. For both noise layers we use the same unraveling, either the $\theta^*$-rotation or the projective one, and we compare the two choices. We distinguish two cases:
\begin{enumerate}
    \item For the $\theta^\ast$-rotation unraveling, the first noise layer uses the Kraus operator $K^\ast_1$. Since the following unitary $U$ is Haar-random and since the state $\ket{\psi'}$ is equally likely to $\ket{\psi}$, as observed above, the state ensemble after applying $U$ is identical to the one found for the projective unraveling when acting with $K_+$ or $K_-$ in the first noise event. Moreover, the probabilities in the two unravelings match up to $\mathcal{O}(p^2)$. Hence, the previous result applies, and the final magic difference is $\mathcal{O}(p^2)$.
    \item For the $\theta^\ast$-rotation unraveling, the first noise layer uses the Kraus operator $K^\ast_0$. This changes the ensemble compared to the projective case, in which $K_\mathrm{id}=\sqrt{1-2p}I$ is picked instead. After the second noise layer, the final states for the two unravelings (conditional on the choice of the previous Kraus operators for the first layer) are given by
    \begin{align}
    \ket{\psi^*_{0,0}} &= U\ket{\psi} - p U(Z_i-z_i)\ket{\psi} - p(Z_j-z_j) U\ket{\psi} + \mathcal{O}(p^2) \qquad \text{with } \, 1-2p-2p z_j+\mathcal{O}(p^2),\nonumber\\
    \ket{\psi^*_{0,+}} &= \frac{(1+Z_j)U\ket{\psi}}{\sqrt{2(1+z_j)}} + \mathcal{O}(p) \qquad \text{with } \, 2p(1+z_j)+\mathcal{O}(p^2),
    \end{align}
    and
    \begin{align}
    \ket{\psi^\mathrm{proj}_{\mathrm{id},\mathrm{id}}} &= U\ket{\psi} \qquad \text{with } \, 1-2p,\nonumber\\
    \ket{\psi^\mathrm{proj}_{\mathrm{id},+}} &= \frac{(1+Z_j)U\ket{\psi}}{\sqrt{2(1+z_j)}} \qquad \text{with } \, p(1+z_j),\\
    \ket{\psi^\mathrm{proj}_{\mathrm{id},-}} &= \frac{(1-Z_j)U\ket{\psi}}{\sqrt{2(1-z_j)}} \qquad \text{with } \, p(1-z_j),\nonumber
    \end{align}
    where we defined $z_i = \bra{\psi}Z_i\ket{\psi}$ and $z_j = \bra{\psi}U^\dagger Z_jU\ket{\psi}$. Repeating the analysis made above, the difference of the SREs between the two unravelings vanishes at $\mathcal{O}(p)$ when the ensemble average is taken, leaving only an $\mathcal{O}(p^2)$ contribution.
\end{enumerate}

The equivalence of $M_2$ between the unravelings up to $\mathcal{O}(p^2)$ thus extends to adding an extra layer of unitaries and noise. This result follows from the fact that the SRE is a smooth functional of the state, and that it is invariant under a bit flip, i.e., $M_2[X_i\ket{\psi}]=M_2[\ket{\psi}]$. The same analysis therefore applies identically to any functional of the state with these properties. In particular, it extends to the SRE evaluated at the end of the circuit, leveraging the Haar-randomness of gates. To conclude the proof for arbitrary circuit depth, let us now consider a mixed protocol with $S=n\cdot N_L$ total noise events in which the projective unraveling is used for the first $0\leq s\leq S$ ones, whereas the $\theta^\ast$-rotation unraveling is implemented for all subsequent ones. Let $M_2(s)$ be the ensemble-averaged SRE found at the end of the circuit within this hybrid scheme. From the previous considerations, we have $M_2(s+1)-M_2(s) = \mathcal{O}(p^2)$. It follows that
\begin{equation}\label{eq:delta_magic}
    |M_2(S)-M_2(0)| = \left|\sum_{s=0}^{S-1}(M_2(s+1)-M_2(s))\right| \leq \sum_{s=0}^{S-1}|M_2(s+1)-M_2(s)| = \mathcal{O}(n N_L p^2).
\end{equation}
The left-hand side of the previous equation is exactly the difference between the two unravelings, and it scales as $p^2$ at fixed size and depth. The bound is of course loose, as the mismatches between $s$ and $s+1$ can have different signs and partially cancel, as well as be damped by the evolution after the $s$th event. Still, this provides an explanation of the numerically observed agreement of Fig.~\ref{fig4}.

The previous derivation applies to $T$-doped Clifford gates as well. In this case, each noise event (on site $i$) is preceded by either a Clifford unitary $U$ or by $T_i U$. The states $\ket{\psi}$ and $X_i\ket{\psi}$ are equally likely if the gates $X_i U$ and $X_i T_i U$ generate the same ensembles as $U$ and $T_i U$, respectively. The first one is trivial, as $X_i U$ remains a Clifford-random gate. It can further be seen that $X_i T_i U \propto T_i(X_i S_i U) = T_i U'$, where $S_i=\mathrm{diag}\{1,i\}$ is a (Clifford) phase gate. Hence, all considerations made for Haar-random circuits remain valid in the doped setup. It is interesting to discuss the bound of Eq.~\eqref{eq:delta_magic} in the purely Clifford case of $q=0$. Here, the projective unraveling has no operation that injects magic, thus yielding $M_2=0$ along all its trajectories. Given that the average non-stabilizerness is observed to grow monotonically with the depth of the circuit, it follows that $M_2(s)\geq M_2(s+1)$, as the left-hand side is produced by an additional noise event compared to the right-hand side. This predicts that magic in the $\theta^\ast$-rotation unraveling will grow as $\sim n N_L p^2$ in the number of layers $N_L$, until reaching saturation for sufficiently large $N_L$. The absence of sign cancellations explains why the two unravelings can yield different magic without contradicting the previous proof. Finally, if we reintroduce a small doping $q\ll 1$ we expect that this difference remains partly visible, which qualitatively explains why the two unravelings do not match for $q=0.05$ in Fig.~\ref{fig6}(d), but restore their agreement at larger values of $q$.

\end{document}